\documentclass[twocolumn]{aastex631}
\usepackage{amsmath}

\newcommand{\COiso}{$^{13}\textrm{CO}$}
\newcommand{\COmain}{$^{12}\textrm{CO}$}
\newcommand{\HTWOO}{H$_{2}\textrm{O}$}
\newcommand{\COTWO}{CO$_{2}$}
\newcommand{\methane}{CH$_{4}$}
\newcommand{\ammonia}{NH$_{3}$}
\newcommand{\enstatite}{MgSiO$_{3}$}

\newcommand{\Mjup}{$M_\mathrm{Jup}$}
\newcommand{\Rjup}{$R_\mathrm{Jup}$}
\newcommand{\Msun}{$M_\odot$}
\newcommand{\Teff}{$T_\mathrm{eff}$}
\newcommand{\kms}{km\ s$^{-1}$}

\newcommand{\leiden}{Leiden Observatory, Leiden University, P.O. Box 9513, 2300 RA, Leiden, The Netherlands}
\newcommand{\caltech}{Department of Astronomy, California Institute of Technology, Pasadena, CA 91125, USA}
\newcommand{\gps}{Division of Geological \& Planetary Sciences, California Institute of Technology, Pasadena, CA 91125, USA}
\newcommand{\warwick}{Department of Physics, University of Warwick, Coventry CV4 7AL, UK}
\newcommand{\warwicks}{Centre for Exoplanets and Habitability, University of Warwick, Gibbet Hill Road, Coventry CV4 7AL, UK}
\newcommand{\ipac}{IPAC, Mail Code 100-22, Caltech, 1200 E. California Boulevard, Pasadena, CA 91125, USA}
\newcommand{\mpia}{Max-Planck-Institut f\"ur Astronomie, K\"onigstuhl 17, 69117 Heidelberg, Germany}
\newcommand{\granada}{Instituto de Astrof{\'i}sica de Andaluc{\'i}a (IAA-CSIC), Glorieta de la Astronom{\'i}a s/n, 18008 Granada, Spain}
\newcommand{\galway}{School of Natural Sciences, Center for Astronomy, University of Galway, Galway, H91 CF50, Ireland}
\newcommand{\carnegiew}{Earth and Planets Laboratory, Carnegie Institution for Science, Washington, DC, 20015}

\shorttitle{YSES 1 b and c with CRIRES+}
\shortauthors{Zhang et al.}

\graphicspath{{./}{figures/}}

\begin{document}

\title{The ESO SupJup Survey III: confirmation of \COiso~in YSES 1 b and atmospheric detection of YSES 1 c with CRIRES$^+$}

\author[0000-0003-0097-4414]{Yapeng Zhang}
\altaffiliation{51 Pegasi b fellow}
\affiliation{\caltech}
\affiliation{\leiden}

\author[0000-0001-9282-9462]{Dar\'io Gonz\'alez Picos}
\affiliation{\leiden}

\author[0000-0003-4760-6168]{Sam de Regt}
\affiliation{\leiden}

\author{Ignas A. G. Snellen}
\affiliation{\leiden}

\author{Siddharth Gandhi}
\affiliation{\warwick}
\affiliation{\warwicks}

\author{Christian Ginski}
\affiliation{\galway}

\author{Aurora Y. Kesseli}
\affiliation{\ipac}

\author{Rico Landman}
\affiliation{\leiden}

\author{Paul Molli\`ere}
\affiliation{\mpia}

\author{Evert Nasedkin}
\affiliation{\mpia}

\author{Alejandro S\'anchez-L\'opez}
\affiliation{\granada}

\author{Tomas Stolker}
\affiliation{\leiden}

\author{Julie Inglis}
\affiliation{\gps}

\author[0000-0002-5375-4725]{Heather A. Knutson}
\affiliation{\gps}

\author{Dimitri Mawet}
\affiliation{\caltech}

\author{Nicole Wallack}
\affiliation{\carnegiew}

\author[0000-0002-6618-1137]{Jerry W. Xuan}
\affiliation{\caltech}

\begin{abstract}

High-resolution spectroscopic characterization of young super-Jovian planets enables 
precise constraints on elemental and isotopic abundances of their atmospheres.
As part of the ESO SupJup Survey, we present high-resolution spectral observations of 
two wide-orbit super-Jupiters in YSES~1 (or TYC 8998-760-1) using the upgraded VLT/CRIRES$^+$ ($\mathcal{R}\sim 100,000$)
in K-band. 
We carry out free atmospheric retrieval analyses to constrain chemical and isotopic abundances, temperature structures, 
rotation velocities ($v\sin i$), and radial velocities (RV). 
We confirm the previous detection of \COiso~in YSES~1~b at a higher significance of 12.6$\sigma$, but point to a higher \COmain/\COiso~ratio of $88\pm 13$ (1$\sigma$ confidence interval), consistent with the primary's isotope ratio $66 \pm 5$. 
We retrieve a solar-like composition in YSES~1~b with a C/O$=0.57 \pm 0.01$,
indicating a formation via gravitational instability or core accretion beyond the CO iceline. 
Additionally, the observations lead to detections of \HTWOO~and CO in the outer planet YSES~1~c at 7.3$\sigma$ and 5.7$\sigma$, respectively.
We constrain the atmospheric C/O ratio of YSES~1~c to be either solar or subsolar (C/O=$0.36 \pm 0.15$), indicating the accretion of oxygen-rich solids.
The two companions have distinct $v\sin i$, $5.34 \pm 0.14$ \kms~for YSES~1~b and $11.3 \pm 2.1$ \kms~for YSES~1~c, despite their similar natal environments.
This may indicate different spin axis inclinations or effective magnetic braking by the long-lived circumplanetary disk around YSES~1~b.
YSES~1 represents an intriguing system for comparative studies of super-Jovian companions and linking present atmospheres to formation histories.

\end{abstract}


\section{Introduction} \label{sec:intro}

Young, self-luminous super-Jovian companions ($<20$ \Mjup) discovered by direct imaging surveys have 
revealed an outstanding population that poses challenges to planet formation theories. 
Typically straddling the mass boundary between planets and brown dwarfs ($\sim$13 \Mjup), they 
orbit their primary stars at large separations. 
Their formation mechanisms remain under debate.
Core accretion in the outer protoplanetary disk is impeded by the long timescale needed to assemble cores massive enough
for the onset of runaway accretion before the gas disk dissipates \citep{LambrechtsJohansen2012}, while
disk instability and cloud fragmentation are expected to form objects with masses in the brown dwarf or stellar regime \citep{ZhuEtAl2012, ForganRice2013a, OffnerEtAl2023}.
Therefore, super-Jovian companions are not easily compatible with formation via either in-situ core accretion or 
gravitational collapse \citep{PollackEtAl1996, Boss1997, Chabrier2003, KratterLodato2016}. 
Alternatively, some of these companions may have formed on closer orbits and then been scattered outward \citep{VerasEtAl2009}, but evidence of such dynamical events has yet to be found, such as excited orbital eccentricity or obliquity of these companions and the potential presence of additional companions with similar or larger masses \citep{BryanEtAl2016, BryanEtAl2020a}. 
Previous studies on orbital architectures of super-Jovian companions versus those of massive brown dwarfs
found evidence for distinct distributions in their semimajor axes and orbital eccentricities 
\citep{NielsenEtAl2019, BowlerEtAl2020, NagpalEtAl2023, DoOEtAl2023}. 
This suggests that the two populations likely have distinct formation mechanisms, but it is unclear
whether a well-defined mass boundary separates them.

Atmospheric characterization provides a critical avenue to distinguish between competing 
formation mechanisms for gas giant planets.
Elemental abundances including the metallicty [M/H] and carbon-to-oxygen (C/O) ratios 
have been suggested as probes for planet formation \citep[e.g. ][]{ObergEtAl2011, Madhusudhan2012, TurriniEtAl2021}.
In general, formation via gravitational instability or cloud fragmentation is thought 
to result in compositions akin to stars,
while core accretion can result in a variety of compositions depending on the 
accretion mechanisms, formation location relative to disk icelines, and migration histories, among other factors 
\citep[e.g.][]{MadhusudhanEtAl2014, MordasiniEtAl2016, BitschEtAl2019}. 
Gravitational instability or cloud fragmentation occurs early when the compositions of disks or clouds are pristine, and the rapid top-down collapses reset the chemical abundances of companions to protostellar values.
As for core accretion, if a companion formed beyond the CO iceline (where most super-Jovian companions are currently located), its atmospheric C/O is expected to be stellar to super-stellar depending on its relative solid versus gas accretion.
If the atmospheric compositions are predominately set by gas accretion, the C/H and O/H are expected to be substellar in the companion's atmosphere 
as most C- and O-bearing volatiles are condensed out in the disk, and the atmospheric C/O is likely super-stellar as the gas disk has an elevated C/O ratio \citep{BerginEtAl2024a}.
On the other hand, if solid accretion plays a significant role (leading to a metal-rich atmosphere), the companion obtains its C and O contents predominately from the ice reservoir, which has a stellar C/O ratio.
\cite{XuanEtAl2024a} studied eight substellar companions with masses of $10-30$ \Mjup,
suggesting that they display broadly solar compositions, which points towards 
formation via gravitational instability.

At an individual level, linking the atmospheric composition of a planet to its formation history 
is further complicated by a chain of processes, including chemistry evolution and pebble drift in disks, 
planet migration, late solid enrichment, and mixing of planet's core and envelope \citep{MolliereEtAl2022a}.
The challenging nature of this task calls for combining different lines of evidence 
from atmospheric tracers such as isotopologue ratios  \citep{MolliereSnellen2019, MorleyEtAl2019} 
and refractory-to-volatile ratios \citep{LothringerEtAl2021, ChachanEtAl2023} 
as well as dynamical indicators 
such as spin and orbital architecture \citep{SnellenEtAl2014, BryanEtAl2018,  BryanEtAl2020a}.
High-resolution spectroscopy of super-Jovian companions provides information on these various aspects simultaneously, 
allowing us to reveal the complete picture of planet formation and evolution history.

Carbon is one of the most accessible elements for probing isotopic composition in exoplanet atmospheres \citep{MolliereSnellen2019}.
\cite{ZhangEtAl2021a} reported the first detection of \COiso~for a super-Jovian companion using 
the medium-resolution integral-field spectrograph SINFONI on the VLT. This study suggested that YSES~1~b's atmosphere has \COiso-that is enriched by a factor of two compared to the interstellar medium, which may 
result from the accretion of \COiso-enriched ices beyond the CO iceline during planet formation. 
However, the link between isotope ratios in protoplanetary disks and planets remains elusive. 
A handful of measurements of carbon isotope abundances in TW Hya disk have been achieved, such as \cite{ZhangEtAl2017, YoshidaEtAl2022, YoshidaEtAl2024, BerginEtAl2024},
indicating strong carbon isotope fractionation in various species.
In contrast, modeling predictions tend to suggest less significant carbon fractionation in disks while finding a
potential \COiso-enrichment in the gas phase \citep{MiotelloEtAl2014, BerginEtAl2024, LeeEtAl2024a}. 
This could potentially result in a deviation from the proto-stellar isotope ratio in planets depending on the gas versus solid accretion.
As carbon isotope ratios are measured in a growing number of exoplanets and brown dwarfs
\citep[e.g.,][]{ZhangEtAl2021, LineEtAl2021, ZhangEtAl2022b, FinnertyEtAl2023, FinnertyEtAl2023a, GandhiEtAl2023, CostesEtAl2024, XuanEtAl2024a, XuanEtAl2024, HoodEtAl2024, LewEtAl2024, deRegtEtAl2024, GonzalezPicosEtAl2024}, which show a variety of \COmain/\COiso~ratios ranging from 30 to 180,
re-observations of the unusually \COiso-enriched or depleted targets with better data quality are needed to confirm the deviation of 
the carbon isotope ratio and assess its potential as a formation tracer.

Super-Jovian companions on wide orbits are excellent targets for directly probing planetary emission at high signal-to-noise, 
therefore allowing for robust constraints on their atmospheric properties and tests on formation hypotheses.
Techniques that combine both the spatial and spectral resolving capability 
have proven to be powerful in characterizing these companions
\citep{SnellenEtAl2014, SnellenEtAl2015, HoeijmakersEtAl2018, RuffioEtAl2023}.
New instruments coupling adaptive optics systems with high-resolution 
spectrographs, such as the Keck Planet Imager and Characterizer \citep[KPIC,][]{MawetEtAl2017}, 
HiRISE on VLT \citep{ViganEtAl2023}, and REACH on Subaru \citep{KotaniEtAl2020a}, 
enhance the contrast limit of detection and high-resolution characterization 
towards more challenging targets \citep[e.g., HR 8799 planets;][]{WangEtAl2021}. 
The high spectral resolving power enables us to better distinguish isotopologues and derive more robust abundance measurements. 
The recently upgraded Cryogenic high-resolution infrared echelle spectrograph \citep[VLT/CRIRES$^+$,][]{DornEtAl2014, DornEtAl2023}
has begun delivering atmospheric characterization of 
directly imaged exoplanets and brown dwarfs \citep{ZhangEtAl2022b, LandmanEtAl2024, ParkerEtAl2024, deRegtEtAl2024, GonzalezPicosEtAl2024}.
Specifically, the ESO SupJup Survey (100-hour large program, ID: 1110.C-4264(F), PI: Snellen) is carrying
out a deep spectroscopic survey on a sample ($\sim$40) of super-Jovian companions and free-floating brown dwarfs
to shed light on the formation pathways of different populations. 
We refer readers to \cite{deRegtEtAl2024} for an overview of the SupJup survey.
More results from the KPIC survey and ESO SupJup survey are forthcoming.

As part of the ESO SupJup survey, this paper focuses on the unique super-Jovian system YSES 1. 
YSES 1 was discovered by the Young Suns Exoplanet Survey 
\citep[YSES,][]{BohnEtAl2020, BohnEtAl2020a}. With two super-Jovian companions, it is the first directly imaged 
multi-planet system around a solar-type star. Comparative studies of multi-planet systems provide an excellent opportunity for testing formation models, as the two planets experienced the same natal environment at different orbital separations
In this paper, we present the atmospheric characterization of YSES~1~b and c
using VLT/CRIRES$^+$ observations.
The paper is organized as follows. In Section~\ref{sec:system}, we introduce the target system
YSES 1. The CRIRES$^+$ observations and data reduction are presented in Section~\ref{sec:observation}.
Then, we fit spectra with disequilibrium and free chemistry retrieval frameworks described in Section~\ref{sec:retrieval}.
We present molecular detections and retrieval results for both planets in Section~\ref{sec:result}.
We compare the CRIRES$^+$ results for YSES~1~b to the previous study with VLT/SINFONI \citep{ZhangEtAl2021a}.
We discuss the rotation velocities, chemical abundances, and their implications for planet formation in Section~\ref{sec:discussion}.
Finally, we summarize our findings in Section~\ref{sec:conclusion}.

\begin{deluxetable*}{lcccc}
    \tablecaption{\label{tab:system}Properties of YSES~1~system}
    \tablehead{\colhead{Property} & \colhead{A} & \colhead{b} & \colhead{c} & \colhead{Reference}}
    \startdata
    \hline
    $\alpha_{2000.0}$ & 13:25:12.13 & & & 1 \\
    $\delta_{2000.0}$ & -64:56:20.69  & & & 1 \\
    Distance (pc) & $ 94.2 \pm 0.1 $  & & & 1 \\
    Age (Myr) & 17 or 27 &  & & 2,3 \\
    $K_s$ (mag) & 8.39 & $\sim$14.7 & $\sim$18 & 2 \\
    Spectral type & K3IV & L0&  L7.5 & 2 \\
    \Teff (K) &  $4573\pm 10$  & $1727^{+172}_{-127}$  & $1240^{+160}_{-170} $ & 2 \\
    Mass & $ 1.00 \pm 0.02 $ $M_\odot$ & $14\pm 3$ \Mjup & $6\pm1$ \Mjup &2 \\
     &  & $21.8\pm3$ \Mjup & $7.2\pm0.7$ \Mjup & 3 \\
     Radius & $1.01\pm 0.02$ $R_\odot$ & $3.0^{+0.2}_{-0.7}$ \Rjup& $1.1^{+0.6}_{-0.3}$ \Rjup &  1,2\\
     $\log g$ (cgs) & $4.43 \pm 0.02$ & $4.10 \pm 0.05$ & $3.3 \pm 0.5$ &  4 \\
     $[\mathrm{Fe/H}]$ & $-0.07\pm0.01$ & $0.04\pm0.05$ & $-0.26\pm0.40$  &  1,4 \\
    RV $\textrm{(\kms)}$  & $12.9 \pm 0.03$  & $11.10 \pm 0.02$ & $13.2\pm1.1$ & 4 \\
    $v\sin i$ $\textrm{(\kms)}$ & $\sim11.1$ & $5.34\pm 0.14$ & $11.3\pm 2.1$ & 4\\
    C/O & - & $0.57 \pm 0.01$ & $0.36\pm0.15$ & 4 \\
    \COmain/\COiso & $66\pm5$ & $88\pm13$&- & 4 \\
    \hline
    \enddata
    \tablerefs{(1) \citet{GaiaCollaborationEtAl2021}, 
    (2) \cite{BohnEtAl2020, BohnEtAl2020a}, (3) \cite{WoodEtAl2023},
    (4) this work.}
\end{deluxetable*}

\section{YSES 1 system} \label{sec:system}

YSES 1, also named TYC 8998-760-1, is a solar analog hosting two planetary-mass companions \citep{BohnEtAl2020, BohnEtAl2020a}.
The primary is a K3IV star with a mass of 1 \Msun.
YSES~1~b and c are located at a projected separation of 160 au and 320 au, 
with an estimated mass of 14$\pm$3 \Mjup~and 6$\pm$1 \Mjup, respectively.
Located at a distance of $94.6\pm0.3$ pc \citep{GaiaCollaborationEtAl2021}, 
YSES 1 was initially classified as a member of the 
Lower-Centaurus Crux (LCC) subgroup of the Scorpius-Centaurus association (Sco-Cen),
with an age of $\sim$17 Myr \citep{PecautMamajek2016}.
However, a recent kinematic study by \cite{WoodEtAl2023} suggested that it is instead a high-probability candidate member of 
the MELANGE-4, a $\sim$27 Myr old extended population of the LCC subgroup. The older age increases the estimated masses of the
companions to 21.8$\pm$3 \Mjup~and 7.2$\pm$0.7 \Mjup~for YSES 1 b and c, respectively \citep{WoodEtAl2023}.
We summarize the properties of the YSES 1 system in Table~\ref{tab:system}.

Based on its photometry from 1 to 5 $\mu$m, the companion YSES~1~b has an effective temperature of 
\Teff$=1727_{-127}^{+172}$ K and a radius of $3.0_{-0.7}^{+0.2}$ \Rjup~\citep{BohnEtAl2020}.
\cite{ZhangEtAl2021a} carried out the spectral characterization of YSES~1~b using the 
medium-resolution ($\mathcal{R}\sim4,500$) integral-field spectrograph SINFONI on the VLT.
The $K$-band spectra led to the first detection of the minor isotopologue \COiso~in exoplanet atmospheres. 
This study reported a solar-like atmospheric C/O ratio of $0.52^{+0.04}_{-0.03}$ and a \COiso-rich
isotopic composition with \COmain/\COiso $=31^{+17}_{-10}$.
In addition, the SINFONI observation revealed Brackett $\gamma$ excess emission, 
indicating active accretion onto YSES~1~b with an estimated accretion rate of 
$10^{-11.1\pm1.6} M_\odot\,yr^{-1}$ \footnote{We note the mistake in the value originally reported in \cite{ZhangEtAl2021a} due to a miscalulation of the Br $\gamma$ emission flux. The accretion rate of YSES~1~b should be lower than previously reported.}. 
With near-infrared polarimetric imaging, \cite{vanHolsteinEtAl2021} found a 3$\sigma$ upper limit 
of 0.6\% on the degree of polarization. This non-detection of polarization suggests that the circumplanetary disk around it 
likely has a low inclination, if present at all.

The companion YSES~1~c is estimated to have a
\Teff$=1240_{-170}^{+160}$ K and a radius of $1.1_{-0.3}^{+0.6}$ \Rjup~using photometric measurements \citep{BohnEtAl2020a}.
Its $H$-$K$ color is the reddest among other directly imaged companions, located close to the L/T transition \citep{BohnEtAl2020a}. Its low surface gravity likely enhances the cloud opacity in its photosphere and causes the red color. 
There is currently no published spectral characterization of companion c.
JWST Cycle 1 program GO 2044 \cite{WilcombEtAl2021} has taken panchromatic spectroscopic observations of the YSES~1 planets 
using NIRSpec and MIRI, which will enable a complete view of this system.

\section{Observations and Data Reduction} \label{sec:observation}

\begin{figure}[t]
    \includegraphics[width=\linewidth]{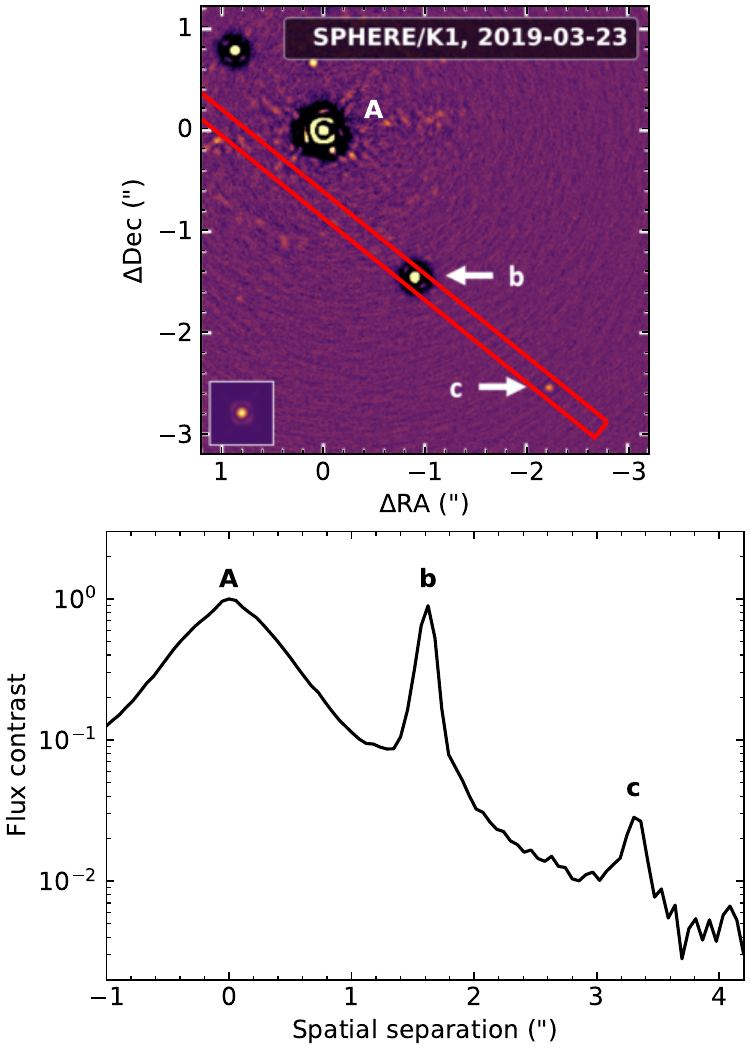}
    \caption{Position of the CRIRES$^+$ slit at one nodding position and wavelength-averaged point spread function (PSF) along the slit in one order (2.32-2.34 \micron) 
    of the CRIRES$^+$ observations. The red box in the upper panel marks the slit position aligning the two YSES 1 companions. The background SPHERE image is adapted from \citep{BohnEtAl2020a}. In the lower panel, the in-slit light from the primary is largely reduced
    because of the slit offset from the stellar center.
    \label{fig:psf}}
\end{figure}

\subsection{CRIRES observations}

We observed YSES 1~b and c with CRIRES$^+$ on UT 2023 February 27 and 28 as part of the ESO SupJup survey 
(Program ID: 1110.C-4264(F), PI: Snellen).
We chose the wavelength setting of K2166 and the slit width of 0.2\arcsec~to cover the
\COiso~bandhead near 2.345 $\mu$m
and achieve the highest spectral resolution ($\mathcal{R}\sim100,000$).
The observations were carried out in AO mode, guided on the primary star. 
We offset the slit to align both companions such that the spectra of YSES~1~b and c 
can be recorded simultaneously. This also allows some off-axis starlight 
to leak into the slit, which provides the stellar spectrum.
The observations were taken using the standard ABBA nodding scheme with a 600-second 
exposure time. We took 22 exposures on the first night and 10 on the second night, 
amounting to 5.3 hours of total integration time.
The seeing was excellent on both nights ($\sim$0.35-0.78\arcsec). 
The target airmass was 1.3-1.4 and 1.4-1.6 for two nights, respectively.
We briefly observed a B0 star tet Mus as a telluric standard star before observing the YSES 1 system.

\subsection{Data Reduction}

We reduced the observations with a customized pipeline \texttt{excalibuhr}\footnote{\url{https://github.com/yapenzhang/excalibuhr}} \citep{ZhangEtAl2024a}.
This end-to-end pipeline can preprocess raw calibration files, including darks, flats,
and lamp frames, trace spectral orders on 2D detector images, apply calibrations to science frames,
remove the sky background by nodding subtraction, combine frames per nodding position, 
extract 1D spectrum, and perform wavelength and flux calibration.
We defer the in-depth description of general pipeline steps to Appendix~\ref{app:pipeline}.

\begin{figure}[t]
    \includegraphics[width=\linewidth]{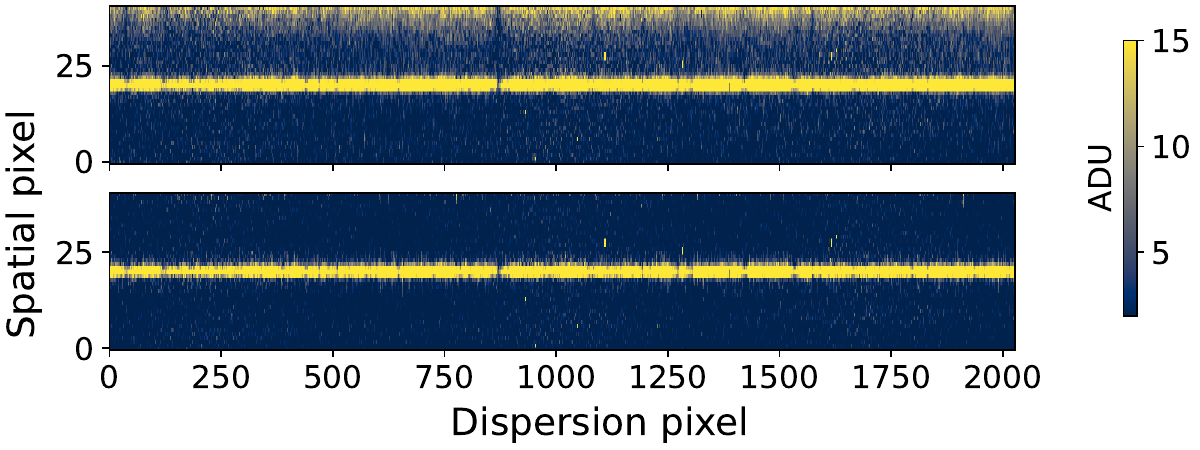}
    \caption{Stellar background removal around the spectral trace of YSES~1~b.
    The upper and lower panels show the cuts of a single order on the detector image 
    before and after removing stellar backgrounds using polynomials.
    \label{fig:remove_bkg}}
\end{figure}

In the observations of YSES 1 b and c, the companions are spatially resolved at separations of
1.6\arcsec~and 3.3\arcsec~from the stellar spectral trace. 
We show the white-light point spread function (PSF) along the slit in Fig.~\ref{fig:psf}.
The starlight contributes roughly $5-15\%$ of the total flux at the separation of companion b. 
To remove the stellar contamination, we carried out additional corrections on the 2D data before spectrum extraction.
Using a fifth-order polynomial, we fit the starlight PSF while excluding 
6 pixels closest to the companion location in spatial directions.  
Then, we evaluated the polynomial on the 
companion position and removed the estimated stellar background from the data.
The choice of the degree of the polynomial was manually optimized to achieve lowest background residuals in the wavelength-averaged PSF. 
Since we perform this correction at each wavelength channel, the wavelength-dependent 
contamination from the star has been appropriately taken into account.
Fig.~\ref{fig:remove_bkg} shows an example of the stellar background correction. 
Similarly, we corrected for the stellar flux near companion c using a third-order polynomial
as the PSF slope of the starlight is smaller for the outer companion.

\begin{figure*}[ht]
    \includegraphics[width=\linewidth]{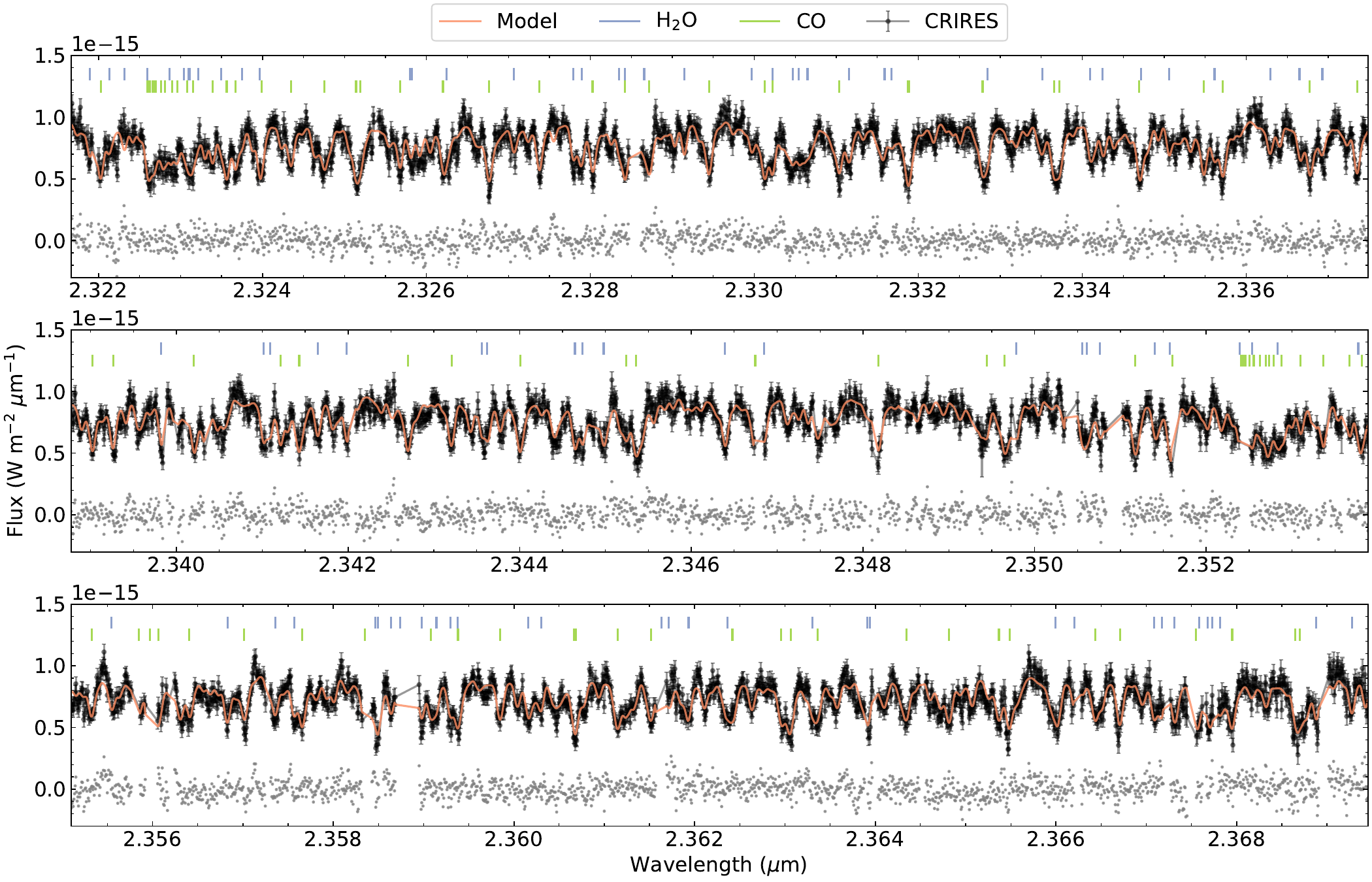}
    \caption{One spectral order of CRIRES$^+$ observations of YSES~1~b from 2.32 to 2.37 \micron. 
    The observations are shown in black data points with error bars. Overplotted is the best-fit model 
    obtained with retrieval analyses.
    We annotated the positions of absorption features from \HTWOO~and CO with short bars on the top of each panel.
    The observational residuals (data minus model) are shown with scattered dots in gray. 
    \label{fig:1b_spec}}
\end{figure*}

Subsequently, we extracted the 1D spectrum of YSES~1~b following the optimal extraction 
method \citep{Horne1986} and calibrated the spectral shape using the early-type 
standard star observations, as explained in Appendix~\ref{app:pipeline}. 
To correct for the telluric absorption lines, we used the ESO sky software \texttt{Molecfit}
\citep{SmetteEtAl2015}, which fits telluric atmospheric transmission model to 
the extracted spectrum of the primary star. We then divided the best-fit telluric model 
from the companion's spectrum to correct for tellurics. The wavelengths affected by
strong telluric absorptions with transmission less than 70\% are masked in our subsequent analysis.  
Finally, the flux was scaled to its $K$-band photometry
of $0.88\times 10^{-15}$ $\mathrm{W\,m}^{-2}\, \mu\mathrm{m}^{-1}$ 
as measured from VLT/SPHERE imaging \citep{BohnEtAl2020}. 
A portion of the spectrum of YSES~1~b is shown in Fig.~\ref{fig:1b_spec}. We reach a S/N of $\sim$15 
per wavelength channel at a spectral resolution $\mathcal{R}\sim$116,000. 
We also extracted the spectrum of the primary YSES~1 and obtained a S/N of $\sim$40 per wavelength channel.

For the low S/N detection of YSES~1~c, we used a simplified 
optimal extraction -- taking the sum of the flux along the spatial direction weighted 
by a Gaussian profile fitted to the white-light PSF (see Fig.~\ref{fig:psf}). 
The sky background noise dominates over the planetary signal in individual wavelength channels. 
Given the low S/N, we discarded the continuum information by high-pass filtering the spectrum 
of YSES~1~c using a 200-pixel ($\sim$2 nm) wide Gaussian filter.

\section{Retrieval Analysis} \label{sec:retrieval}

\subsection{Spectral Model} \label{sec:forward}

The spectral model of the companion consists of three components: the temperature
model, the chemistry model, and the cloud model. The models are set up as follows. 

\subsubsection{Temperature Model} \label{sec:forward-tp}

To model the temperature structure of YSES~1~b, 
we parametrize the temperature-pressure (T-P) profile following \cite{ZhangEtAl2023}, 
using temperature gradients 
$\Delta \ln T/ \Delta \ln P$ across 6 pressure knots from $10^{-5}$ to 10 bar 
and one absolute temperature value $T_0$ at 10 bar.
The pressure knots are set at $10^{-5}$, $10^{-3.5}$, $10^{-2}$, $10^{-1}$, $10^{0}$, and $10^{1}$ bar.
The whole T-P profile is determined by a spline interpolation onto 60 atmospheric 
layers evenly spaced in log pressure between $10^{-5}$ and 10 bar.
The choice of priors for the temperature gradients is informed by the T-P profiles from 
the self-consistent Sonora Bobcat model grids \citep{MarleyEtAl2021}. 
We calculate the temperature gradients of Sonora profiles with \Teff=1500-2500 K and 
take a uniform prior within the 3$\sigma$ values (see Table~\ref{tab:1b_params}). 

For the low S/N observations on YSES~1~c, we choose to reduce the complexity of the T-P model
through linear interpolation across precomputed T-P profiles from Sonora Bobcat model grids.
We adopt the T-P model grid with solar metallicity and C/O ratio.
This model then only involves two free parameters: the surface gravity and effective temperature.

\subsubsection{Chemistry Model}

We explore two different chemistry models: disequilibrium chemistry and free chemistry.
The disequilibrium chemistry model first assumes chemical equilibrium to determine 
the chemical abundances at a given pressure, temperature, carbon-to-oxygen ratio (C/O), 
and metallicity ([M/H]) by interpolation of a precomputed table \citep{MolliereEtAl2017, MolliereEtAl2020}. 
The model parameterizes the disequilibrium chemistry of major species (CO, \methane, and \HTWOO) by enforcing constant 
mass fractions above a certain quenching pressure level ($P_\mathrm{quench}$), 
which approximates the effect of vertical mixing in disequilibrium chemistry \citep{ZahnleMarley2014}.
The free chemistry model allows the abundance of each chemical species to vary while assuming a vertically constant profile.
In the case of chemical quenching at a deep atmospheric region below the photosphere, 
this constant abundance profile can be a reasonable approximation, which is expected to 
lead to similar retrieval results as the disequilibrium chemistry model.

\subsubsection{Cloud Model}

We adopt the condensate cloud model from \citet{AckermanMarley2001} 
with silicate or Fe as cloud opacity sources, which are expected to be the dominant cloud 
species in L dwarfs \citep{CushingEtAl2006}. 
The cloud is characterized by four parameters: the mass fraction 
of the cloud species at the cloud base $X_0^\mathrm{species}$, the settling parameter $f_\mathrm{sed}$ 
(controlling the thickness of the cloud above the cloud base), 
the vertical eddy diffusion coefficient $K_\mathrm{zz}$ (effectively determining the particle size), 
and the width of the log-normal particle size distribution $\sigma_g$.
Following \cite{MolliereEtAl2020}, the location of the cloud base $P_\mathrm{base}$ is 
determined by intersecting the condensation curve of the cloud species with the T-P 
profile of the atmosphere.
For the relatively hot atmosphere of YSES~1~b, we adopt Fe as the cloud species as it is expected to condense out near the photosphere. For the cooler atmosphere of YSES~1~c, \enstatite~cloud is adopted. We note that the specific choice of cloud species has little influence on the retrieval results for the K-band high-resolution spectra in our test runs.

\subsection{Retrieval Framework}

Given the model setup detailed in Section~\ref{sec:forward}, we compute synthetic spectra 
of the companion using the radiative transfer code \texttt{petitRADTRANS} \citep[\texttt{pRT},][]{MolliereEtAl2019}. 
The model accounts for the Rayleigh scattering of H$_2$ and He, the collision-induced absorption of H$_2$-H$_2$ and H$_2$-He, and
the scattering and absorption cross-sections of crystalline, irregularly shaped cloud particles. 
We use the line-by-line mode of \texttt{pRT} to calculate the emission spectra at high spectral resolution. 
To speed up the calculation, we downsample the original opacity tables (with $\lambda/\Delta\lambda \sim10^6$) by a factor of 2. 
We benchmarked the downsampling factor with test runs to ensure that it does not bias the retrieval results \citep[see also ][]{ZhangEtAl2021a, XuanEtAl2022}.
We include opacity from \HTWOO~\citep{PolyanskyEtAl2018}, \COmain, \COiso~\citep{LiEtAl2015}, \methane~\citep{HargreavesEtAl2020}, 
\COTWO~\citep{RothmanEtAl2010}, \ammonia~\citep{ColesEtAl2019}, H$_2$S \citep{AzzamEtAl2016}, and HF \citep{CoxonHajigeorgiou2015} in our model.

Subsequently, the synthetic high-resolution spectrum is radial-velocity (RV) shifted 
by the systemic and barycentric velocity, and rotationally broadened by $v\sin i$ using the method from
\cite{CarvalhoJohns-Krull2023}.
The model spectrum at native resolution is then convolved with a Voigt kernel, 
a combination of Gaussian and Lorentzian kernels, 
to match the resolving power of the instrument ($\lambda/\Delta\lambda \sim100,000$).
The width of the Gaussian kernel and the Lorentzian parameter are determined to be 2.7 and 0.5 pixels
while fitting the telluric transmission in stellar spectrum using \texttt{Molecfit}.
The spectrum is then binned to the wavelength grid of the observed spectrum and 
scaled by the dilution factor $(R_p/D)^2$ to match the observed flux at Earth.
To account for the imperfect calibration of the broadband spectral shape and the potential extinction due to a circumplanetary disk, we allow for relative flux offsets across the different orders 
while keeping the flux in the central order ($\lambda\sim$2.24-2.26 \micron) fixed.
The offset factor in each order is calculated on-the-fly
by least-square fitting the model spectrum to the observation 
\citep[see also][]{RuffioEtAl2019, LandmanEtAl2024, deRegtEtAl2024}. 
We also run test retrievals allowing for an arbitrary scaling of the absolute flux, which marginalizes the possible inaccuracy of the photometric measurement and the extinction effect, and note that this results in no difference in retrieved parameters. 
For the spectral model of YSES~1~c, we discard the continuum information by 
high-pass filtering of both the observation and model using a 200-pixel ($\sim$2 nm) wide Gaussian filter. 
As no absolute flux information is associated with the data, 
we add a free scaling parameter $f_\mathrm{scaling}$ for each night to match the line strengths of models and data.

The log-likelihood of the model $\mathbf{M}$ given the observation $\mathbf{d}$ is formulated as:
\begin{equation}
    \begin{aligned}
        \ln \mathcal{L} = & -\frac{1}{2} \bigg[ N\ln(2\pi)+\ln(|\Sigma_0|)+N\ln(s^2) \\
        & + \frac{1}{s^2}(\mathbf{d} - \mathbf{M})^T \Sigma_0^{-1} (\mathbf{d} - \mathbf{M}) \bigg],
    \end{aligned}
\end{equation}
where $N$ is the number of data points, $\Sigma_0$ is the covariance matrix with the diagonal items populated by observation uncertainties,
and $s$ is the errorbar inflation parameter 
accounting for the underestimated uncertainties. Following \cite{RuffioEtAl2019}, 
the optimal $s$ for each model can be solved by scaling the reduced $\chi^2$ to 1:
\begin{equation}
    s^2 = \frac{1}{N} (\mathbf{d} - \mathbf{M})^T \Sigma_0^{-1} (\mathbf{d} - \mathbf{M}).
\end{equation}
The log-likelihood is evaluated in each spectral order in individual nights before being combined, 
allowing for different linear coefficients and errorbar scaling factors optimized for each order.

In total, the retrieval models have 15 to 20 free parameters, as summarized in Table~\ref{tab:1b_params} and~\ref{tab:1c_params}.
We use the nested sampling tool \texttt{PyMultiNest} \citep{BuchnerEtAl2014}, 
which is a Python wrapper of the \texttt{MultiNest} method \citep{FerozEtAl2009} for the Bayesian inference.
The retrievals are performed in importance nested sampling mode with a constant efficiency of 5\%. 
It uses 1000 live points to sample the parameter space and derives the posterior distribution of free
parameters.

\begin{deluxetable*}{ccccc}
    \tablecaption{Priors and posteriors of YSES~1~b retrievals.}
    \tablehead{\colhead{Parameter} & \colhead{Prior}  & \colhead{Disequilibrium + GP} & \colhead{Disequilibrium} & \colhead{Free} }
    \startdata
    $\log g$ (cm\,s$^{-2}$) & $\mathcal{U}$(3.0, 5.5) & $4.10 \pm 0.05$ & $4.33 \pm 0.05$ & $4.25 \pm 0.05$\\
    $R_p$ ($R_\mathrm{Jup}$) & $\mathcal{U}$(1.0,\ 3.5) & $2.30\pm0.02$ & $2.27\pm0.02$ & $2.27\pm0.02$ \\
    $v\sin i$ (\kms) & $\mathcal{U}$(0,\ 20) & $5.34\pm 0.12$ & $5.34\pm 0.14$ & $5.32\pm 0.13$\\
    RV (\kms) & $\mathcal{U}$(-20,\ 20) & $11.07 \pm 0.03$ & $11.10 \pm 0.02$ & $11.10 \pm 0.02$\\
    $\epsilon_\mathrm{limb}$ & $\mathcal{U}$(0,\ 1) & $0.32 \pm 0.19$ & $0.33 \pm 0.24$ & $0.30 \pm 0.24$\\
    $\rm [M/H]$ & $\mathcal{U}$(-1.5,\ 1.5) & $0.04\pm 0.05$ & $0.20\pm 0.05$ & $0.11\pm 0.05$ \\
    C/O & $\mathcal{U}$(0.1,\ 1.5) & $0.58\pm0.01$ & $0.57\pm0.01$ & $0.61\pm0.01$\\
    log$X^\mathrm{H_2O}$ & $\mathcal{U}$(-12,\ -1) & -& - & $-2.53 \pm 0.05$ \\
    log$X^\mathrm{^{12}CO}$ & $\mathcal{U}$(-12,\ -1) & -& - & $-2.14 \pm 0.05$ \\
    log$X^\mathrm{^{13}CO}$ & $\mathcal{U}$(-12,\ -1) & -& - & $-4.05 \pm 0.07$ \\
    log$X^\mathrm{CH_4}$ & $\mathcal{U}$(-12,\ -1) & - & -& $<-5.9$ \\
    log$X^\mathrm{NH_3}$ & $\mathcal{U}$(-12,\ -1) & - & -&  $<-5.6$\\
    log$X^\mathrm{CO_2}$ & $\mathcal{U}$(-12,\ -1) & - & -&  $<-3.3$\\
    log$X^\mathrm{H_2S}$ & $\mathcal{U}$(-12,\ -1) & - & -&  $<-3.8$\\
    log$X^\mathrm{HF}$ & $\mathcal{U}$(-12,\ -1) & - & -&  $-6.52\pm0.09$\\
    log$P_\mathrm{quench}$ (bar) & $\mathcal{U}$(-5,\ 1) & $<0.8$ & $<0.8$ & - \\
    log(\COmain/\COiso) & $\mathcal{U}$(-12,\ -1) & $-1.95\pm 0.06$ & $-1.91\pm 0.05$ & $-1.93\pm 0.05$ \\
    $T_0$ (K) & $\mathcal{U}$(2000,\ 5000) & $2919 \pm 63$ & $2489 \pm 30$ & $2522 \pm 35$\\
    $(d\ln T /d\ln P)_0$  & $\mathcal{U}$(0.01,\ 0.05) & $0.029 \pm 0.011$ & $0.031 \pm 0.012$ & $0.031 \pm 0.012$ \\
    $(d\ln T /d\ln P)_1$  & $\mathcal{U}$(0.01,\ 0.05)  & $0.033 \pm 0.014$ & $0.021 \pm 0.010$ & $0.021 \pm 0.010$ \\
    $(d\ln T /d\ln P)_2$  & $\mathcal{U}$(0.01,\ 0.25) & $0.012 \pm 0.002$ & $0.011 \pm 0.001$ & $0.011 \pm 0.001$ \\
    $(d\ln T /d\ln P)_3$  & $\mathcal{U}$(0,\ 0.41) & $0.081 \pm 0.003$ & $0.079 \pm 0.002$ &  $0.077 \pm 0.002$ \\
    $(d\ln T /d\ln P)_4$  & $\mathcal{U}$(0.,\ 0.38) & $0.078 \pm 0.004$ & $0.070 \pm 0.003$ & $0.075 \pm 0.004$ \\
    $(d\ln T /d\ln P)_5$  & $\mathcal{U}$(0.07,\ 0.3)  & $0.22 \pm 0.028$  & $0.078 \pm 0.007$ & $0.079 \pm 0.009$ \\
    \Teff (K) & - & $1909 \pm 10$ & $1893 \pm 10$ & -\\
    log($X_0^\mathrm{Fe}/X_\mathrm{eq}^\mathrm{Fe}$)& $\mathcal{U}$(-2,\ 1) & $-0.7\pm 0.8$  & $-1.2\pm 0.8$ & - \\
    log($X_0^\mathrm{Fe}$)& $\mathcal{U}$(-12,\ -1) & - & - & $-4.9\pm 0.6$ \\
    $f_\mathrm{sed}$ & $\mathcal{U}$(0,\ 10) & $5.6 \pm 2.1$  & $6.2 \pm 2.5$ & $5.9 \pm 2.5$\\ 
    log($K_\mathrm{zz}$) & $\mathcal{U}$(5,\ 13) & $2.6 \pm 0.4$ &  $2.9 \pm 3.5$ & $6.0 \pm 2.5$ \\
    $\sigma_g$ & $\mathcal{U}$(1.05,\ 3) & $1.7 \pm 0.5$  & $1.8 \pm 0.6$ & $2.0 \pm 0.6$\\
    \enddata
    \label{tab:1b_params}
    \tablecomments{$\mathcal{U}$(a, b) represents a uniform distribution and 
    $\mathcal{N}$(a, b) represents a normal distribution. The last two columns show
    posteriors with 1$\sigma$ uncertainties from the disequilibrium chemistry model and
    free chemistry model. For unbounded parameters, we show 3$\sigma$ upper limits instead.}
\end{deluxetable*}

\section{Retrieval results} \label{sec:result}

\subsection{YSES~1~b}

Our baseline model uses the free T-P profile, disequilibrium chemistry model, 
and condensate cloud model as detailed in Section~\ref{sec:forward}. 
The posterior distributions of free parameters are summarized in Table~\ref{tab:1b_params}.

\begin{figure*}[t!]
    \includegraphics[width=\linewidth]{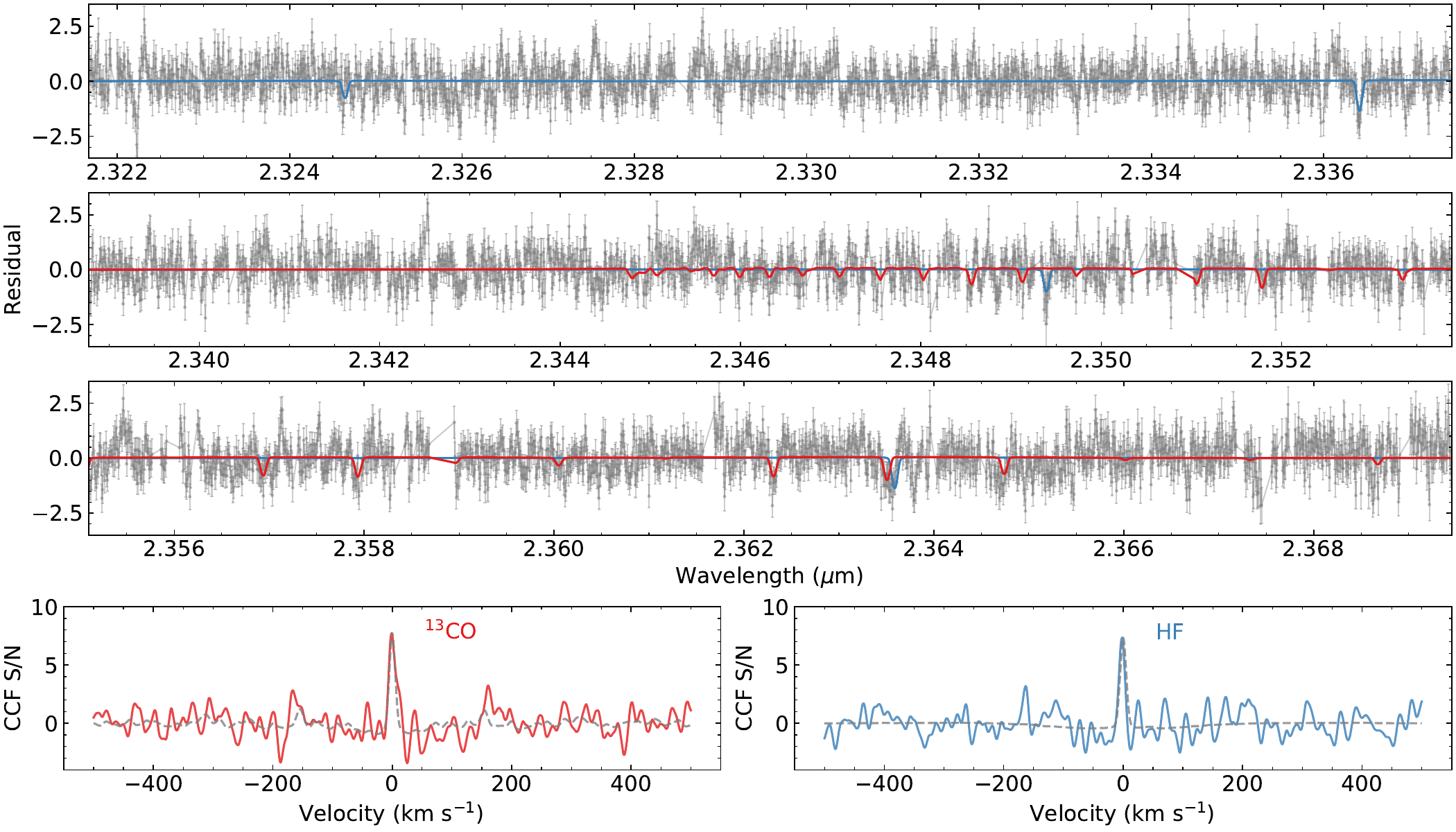}
    \caption{Detection of \COiso~and HF molecules in CRIRES$^+$ observations of YSES~1~b.
      The top three panels show the 2.32-2.37 \micron~observational residuals, where the main \HTWOO~and \COmain~spectral features have been removed from the data.
      The best-fit \COiso~and HF model spectra are overplotted in red and blue lines, demonstrating the detection of individual lines in the data.
      The bottom panels display cross-correlation functions (CCF) of observational residuals with \COiso~and HF models. 
      The gray dashed line is each molecular model's auto-correlation function (ACF).
    \label{fig:1b_ccf}}
\end{figure*}

\subsubsection{Chemistry}

We show the best-fit model spectra overplotted with CRIRES$^+$ observations in Fig.~\ref{fig:1b_spec}.
As annotated in the plot, the \HTWOO~and \COmain~features dominate the K-band spectrum.
We confirm the detection of \COiso~in YSES~1~b, which was initially discovered using 
medium-resolution spectroscopy with VLT/SINFONI \citep{ZhangEtAl2021a}. 
To show the \COiso~detection, we carry out cross-correlation analysis as follows.
The cross-correlation function (CCF) is calculated as 
\begin{equation}
    \mathrm{CCF}(v) = \frac{1}{s^2} \mathbf{M}_\mathrm{sp} (v) ^T  \Sigma_0^{-1} \mathbf{R}.
\end{equation}
where $R$ is the observed spectra minus the best-fit model with the abundance of 
\COiso~being set to zero; $\mathbf{M}_\mathrm{sp}$ is the molecular template computed 
by differencing the best-fit model and the same model without the contribution of that molecule;
$v$ is the radial velocity shift between the model and data.
Both $\mathbf{R}$ and $\mathbf{M}_\mathrm{sp}$ were high-pass filtered using a Gaussian filter with a width of 200 pixels.
Then, the CCFs of individual orders were combined into a master CCF, as shown in Fig.~\ref{fig:1b_ccf}.
The noise of the CCF was estimated by subtracting the model's auto-correlation function (ACF) 
and taking the standard deviation at $|v|>150$ \kms.
The S/N of the \COiso~CCF detection is 7.5, and the individual absorption lines can be identified in the 
observational residuals shown in Fig.~\ref{fig:1b_ccf} thanks to the high S/N and spectral resolution.
To quantify the detection significance, we compare the retrieval models' Bayesian evidence ($Z$) 
while leaving out the \COiso~molecule from opacity sources.
The difference of $\ln Z$ between the models is 76.2, which translates to 12.6$\sigma$ significance 
following \cite{BennekeSeager2013}.
We caution that running \texttt{PyMultinest} in constant sampling efficiency mode may lead to overconfident posterior constraints and Bayesian evidence \citep[see also the discussion in ][]{ChubbMin2022}.
We also ran retrievals with line opacities from H$_2^{18}$O \citep{Polyansky2017} and C$^{18}$O \citep{LiEtAl2015} and found no constraints on the oxygen isotope ratio.

The presence of HF has recently been identified in brown dwarfs in the SupJup Survey and will be discussed in more detail in de Regt et al. (in prep.). HF has then been detected in a growing number of substellar objects \citep{GonzalezPicosEtAl2024}. 
With the free chemistry model, we also detect HF in YSES~1~b at S/N$\sim$7 (see Fig.~\ref{fig:1b_ccf}) 
or 11.7$\sigma$ significance in terms of Bayesian evidence.
As for other molecular species, we found no detection of \methane, \ammonia, \COTWO, and H$_2$S.
The 3$\sigma$ upper limits of their mass fractions are shown in Table~\ref{tab:1b_params}. 
We found no evidence of disequilibrium chemistry for YSES~1~b through the unconstrained quench pressure.
Its atmospheric chemistry is likely in equilibrium because of the relatively high temperature and the active accretion, which could potentially heat the top atmosphere, reduce the vertical temperature gradient and suppress convection.

\begin{figure*}[t]
    \centering
    \includegraphics[width=0.9\linewidth]{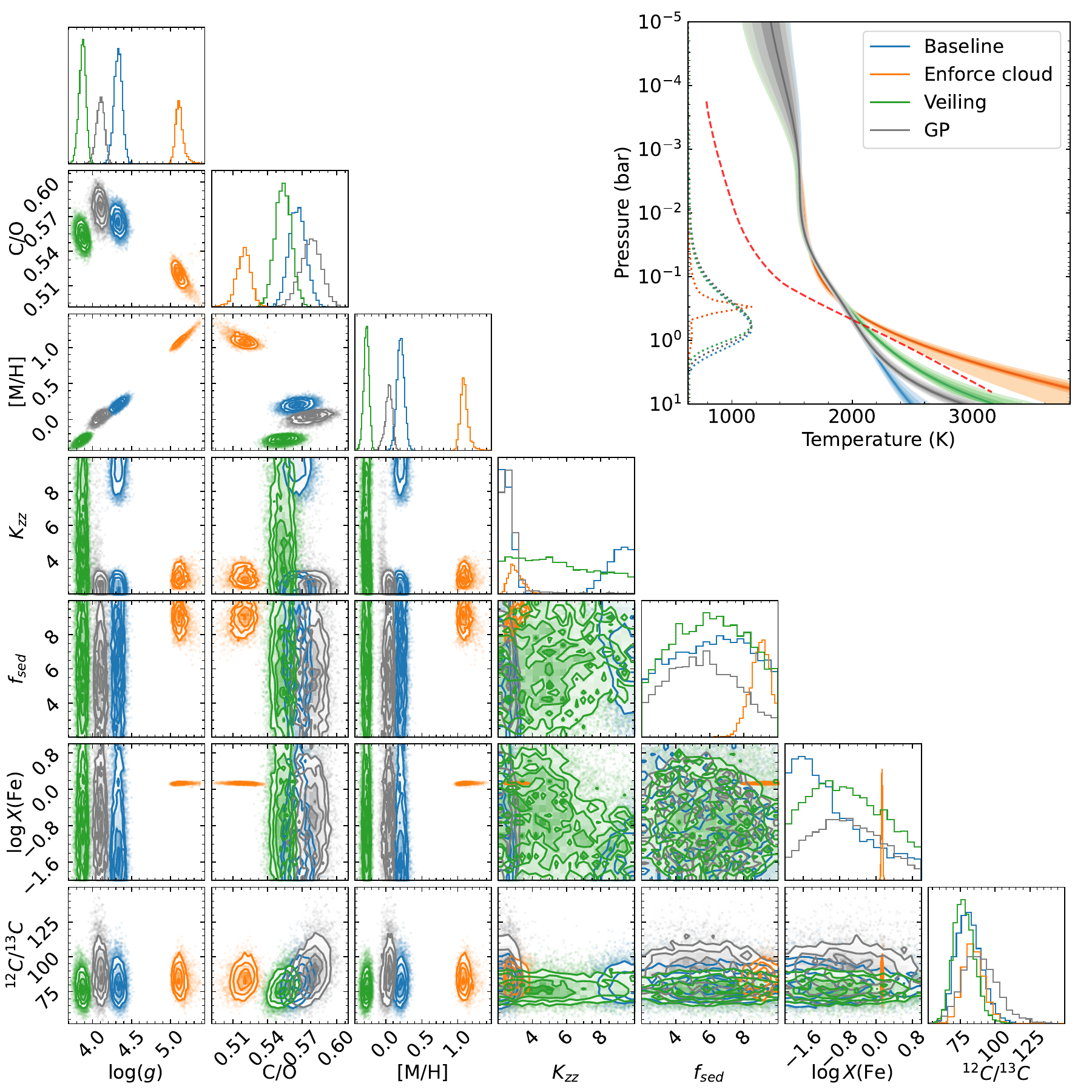}
    \caption{Comparison of retrieval results of alternative models for YSES~1~b.
    Blue represents the baseline model, orange shows the enforced (optically thick) cloud model, green represents the model with veiling effect, and gray is the model with Gaussian Processes accounting for correlated noise.
    The upper right panel shows the retrieved T-P profiles. Different color saturations indicate the envelopes of 
    1$\sigma$, 2$\sigma$, and 3$\sigma$ intervels. The red dashed line is the self-consistent T-P profile (\Teff=1900 K, $\log g$=4.0)
    from the Sonora Bobcat model grid. The dotted curve shows the flux-weighted emission contribution of each model.
    The corner plots show the posteriors of free parameters, including surface gravity, C/O ratio, 
    metallicity, vertical mixing parameter, settling parameter, mass fraction of clouds at the cloud base, and carbon isotope ratio.
    We note the correlation between surface gravity and metallicity, while C/O and \COmain/\COiso~are less model dependent.
    }\label{fig:compare_pt}
\end{figure*}

To assess the effect of correlated noise on the retrieval results, we add Gaussian processes (GP) to the retrieval model 
following \cite{deRegtEtAl2024}. 
The results are consistent with the baseline model without GP, while increasing the uncertainties of the retrieved parameters by $\sim$50\%
as shown in Fig.\ref{fig:compare_pt} and Table~\ref{tab:1b_params}.
In specific, the GP model results in a carbon isotope ratio \COmain/\COiso$=88\pm13$ (1$\sigma$ uncertainty intervals), consistent with the \COmain/\COiso$=81\pm9$ in the baseline model.
This value is consistent with the solar system value of $\sim$89 \citep{AndersGrevesse1989} and the ISM value of $68\pm15$ 
\citep{MilamEtAl2005}, but disagrees with our previous SINFONI result of $31_{-10}^{+17}$ \citep{ZhangEtAl2021a} by 2.7$\sigma$. 
We further discuss this discrepancy in Section~\ref{sec:sinfoni}.
For elemental abundance ratios, we retrieved C/O=$0.57\pm0.01$ and [M/H]=$0.20\pm 0.05$ (1$\sigma$ uncertainty intervals). 
We note that these uncertainties are likely underestimated because they do not account for model uncertainties.
The metallicity, in particular, is usually degenerate with surface gravity \citep{GonzalezPicosEtAl2024} 
and can be highly model-dependent given different clouds and T-P models (see \ref{sec:TP}).

\subsection{T-P profile and clouds} \label{sec:TP}

The retrievals result in T-P profiles that are more isothermal than self-consistent 
model predictions as shown in Fig.\ref{fig:compare_pt}. 
Our derived effective temperature of $\sim$1900 K is consistent with the photometric estimation (see Table \ref{tab:system}) within 1$\sigma$.
Enforcing a self-consistent T-P profiles in the retrieval leads to a significantly worse fit to the data.
Additionally, we test retrievals while removing the continuum of the spectra.
This results in an even more isothermal T-P profile. 
This is a common issue that has been noted in many other free retrieval analyses 
\citep[e.g.,][]{BurninghamEtAl2017, MolliereEtAl2020, BurninghamEtAl2021, ZhangEtAl2021}.
One of the potential explanations is the degeneracy between the free T-P profile and clouds.
Since thick clouds can block the emission from high-temperature regions of the atmosphere below, 
a cloudy atmosphere is spectrally equivalent to a cloudless atmosphere combined 
with a shallow (isothermal) temperature gradient.
In our baseline model, the cloud properties are not well-constrained, and the data are 
consistent with optically thin clouds or cloud-free atmospheres.
We test the effect of trade-offs between clouds and T-P profiles by enforcing optically thick clouds in retrievals following \cite{ZhangEtAl2021}. 
We modify our condensate cloud model by setting the optical depth of the cloud
at the photosphere as a free parameter $\tau_\mathrm{cloud}$ and ensuring $\tau_\mathrm{cloud} > 1$.
The enforced cloud model converges to cloudy solutions, 
resulting in steeper T-P profiles at the bottom of the atmosphere than the baseline model.
The inferred temperature structure in the upper atmosphere barely changes, 
and is still shallower than the prediction of the self-consistent model. 
However, the metallicities in these enforced models increase to more than ten times the solar value,
and the $\log g$ up to 5.2 (see Fig.~\ref{fig:compare_pt}). 
Considering the companion's mass of $14-22$ \Mjup, these bulk properties are not physically preferred solutions.
Therefore, it is unclear if these modified models represent the truth better. 
That the [M/H] and $\log g$ are strongly correlated and model-dependent is commonly found in free 
retrieval analysis \citep{MolliereEtAl2020, ZhangEtAl2021a, GonzalezPicosEtAl2024}. An analytical explanation for this can be found in \cite{MolliereEtAl2015} in Section 5.3-5.5.

In addition to the cloud and T-P trade-offs, spectral inference of young super-Jovian companions can 
sometimes be complicated by their ongoing accretion.
On the one hand, the accretion may result in the veiling of planetary spectra due to the
emission from accretion shock and circumplanetary disk.
The veiling effectively leads to shallower spectral lines, making the companion appear to have an isothermal 
T-P profile at low pressures. 
On the other hand, the isothermal T-P profile may be physically feasible 
as a result of the accretion heating of the atmosphere.
YSES~1~b shows the signature of active accretion in its Brackett $\gamma$ emission \citep{ZhangEtAl2021a}.
Using the analytical model in \cite{Zhu2015}, we estimate the photospheric temperature heated by the 
magnetospheric accretion to be $\sim$2000 K assuming an accretion rate $\dot M\sim 10^{-11} M_\odot\,\mathrm{yr}^{-1}$, 
a typical magnetic field strength $B\sim$100 G, and a filling factor of the accretion column $f\sim$0.01. 
This is similar to the retrieved temperatures of the companion.
However, it is yet unclear how the accretion heating alters the vertical temperature structure.

To assess the potential effect of veiling on retrievals, we parameterize the veiling as a constant flux 
added to the continuum in each order \citep{Sullivan2019}:
\begin{equation}
    F' = \frac{F + r_kF_0}{1 + r_k},
\end{equation}
where $r_k$ is the veiling factor representing the flux ratio of the additional sources to photospheric emission, 
$F$ and $F'$ are the model spectra before and after veiling, 
and $F_0$ is the median flux level of the spectrum $F$.
This veiling model adds five free parameters (corresponding to five spectral orders) to the retrieval. 
The results of the veiling model are compared to the baseline model in Fig.~\ref{fig:compare_pt}.
It shows that the veiling alters the temperature profile similarly to the enforced cloud model 
while leading to significantly different [M/H] and $\log g$.
Given the model-dependency of the metallicity and surface gravity, we cannot constrain them with this observation. 
Regardless, the C/O and \COmain/\COiso~ratios remain robust across different models.

\subsection{YSES~1~c}


\begin{figure*}[t!]
    \includegraphics[width=\linewidth]{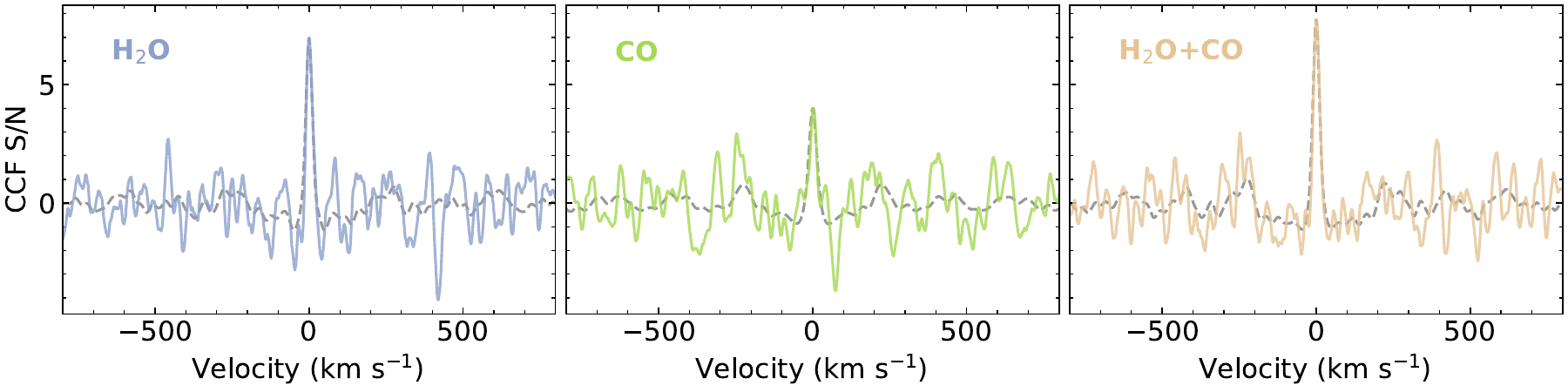}
    \caption{Cross-correlation dectection of \HTWOO~and CO in CRIRES$^+$ observations of YSES~1~c.
    The detection S/N of \HTWOO, CO, and combined models are 7, 4, and 7.8, respectively.
    The gray dashed line shows the auto-correlation function (ACF) of each molecular template.
    \label{fig:1c_ccf}}
\end{figure*}

\begin{deluxetable}{cccc}
    \tablecaption{Priors and posteriors of YSES~1~c retrievals.}
    \label{tab:1c_params}
    \tablehead{\colhead{Parameter} & \colhead{Prior} & \colhead{Disequilibrium} & \colhead{Free}}
    \startdata
    $\log g$ (cm\,s$^{-2}$) & $\mathcal{U}$(3.0, 5.5) & $3.33 \pm 0.36$ & $3.48\pm0.44$ \\
    $v\sin i$ (\kms) & $\mathcal{U}$(0,\ 20) & $11.3\pm 2.1$ & $10.8\pm 2.1$\\
    RV (\kms) & $\mathcal{U}$(-20,\ 20) & $13.3\pm1.1$ & $13.1 \pm 1.1$ \\
    $\epsilon_\mathrm{limb}$ & $\mathcal{U}$(0,\ 1)& $0.52 \pm 0.33$ &$0.52 \pm 0.32$ \\
    \Teff (K) & $\mathcal{U}$(500,\ 2000) & $958 \pm 119$ & $918 \pm 155$\\
    $\rm [M/H]$ & $\mathcal{U}$(-1.5,\ 1.5) & $-0.26\pm 0.40$ & $-0.50\pm 0.36$ \\
    C/O & $\mathcal{U}$(0.1,\ 1.5) & $0.36\pm0.14$ & $0.24\pm0.15$\\
    log$X^\mathrm{H_2O}$ & $\mathcal{U}$(-12,\ -1) & - & $-2.56 \pm 0.32$ \\
    log$X^\mathrm{CO}$ & $\mathcal{U}$(-12,\ -1) & - & $-2.80 \pm 0.37$ \\
    log$X^\mathrm{CH_4}$ & $\mathcal{U}$(-12,\ -1) & - & $<-4.2$ \\
    log$X^\mathrm{NH_3}$ & $\mathcal{U}$(-12,\ -1) & - &  $<-3.9$\\
    log$P_\mathrm{quench}$ (bar) & $\mathcal{U}$(-5,\ 1)  & $0.72\pm0.26$ & - \\
    log($X_0^\mathrm{MgSiO_3}/X_\mathrm{eq}$)& $\mathcal{U}$(-2,\ 1) & $-0.35\pm 0.98$ & - \\
    log($X_0^\mathrm{MgSiO_3}$)& $\mathcal{U}$(-12,\ -1) & - & $-3.2\pm 1.1$ \\
    $f_\mathrm{sed}$ & $\mathcal{U}$(0,\ 10) & $5.6 \pm 2.7$ & $5.8 \pm 2.6$\\ 
    log($K_\mathrm{zz}$) & $\mathcal{U}$(5,\ 13) & $6.0 \pm 2.5$ & $6.0 \pm 2.5$ \\
    $\sigma_g$ & $\mathcal{U}$(1.05,\ 3) & $2.0 \pm 0.7$ & $2.0 \pm 0.6$\\
    log($f_\mathrm{scaling1}$)  & $\mathcal{U}$(-2,1) & $-0.94\pm0.13$ & $-0.93\pm0.12$ \\
    log($f_\mathrm{scaling2}$)  & $\mathcal{U}$(-2,1) & $0.09\pm0.12$ & $0.09\pm0.12$ \\
    \enddata
\end{deluxetable}

The YSES~1~c spectrum has a much lower S/N ($<0.1$ per wavelength channel) due to its relative faintness as shown in Fig.~\ref{fig:1c_spec} in the Appendix. We retrieve its atmospheric properties using the grid 
T-P profile, disequilibrium chemistry (or free chemistry), and condensate cloud model as described in Section~\ref{sec:forward}. 
The posterior distributions of free parameters are summarized in Table~\ref{tab:1c_params}.

\subsubsection{Chemistry}

\begin{figure}[t]
    \centering
    \includegraphics[width=\linewidth]{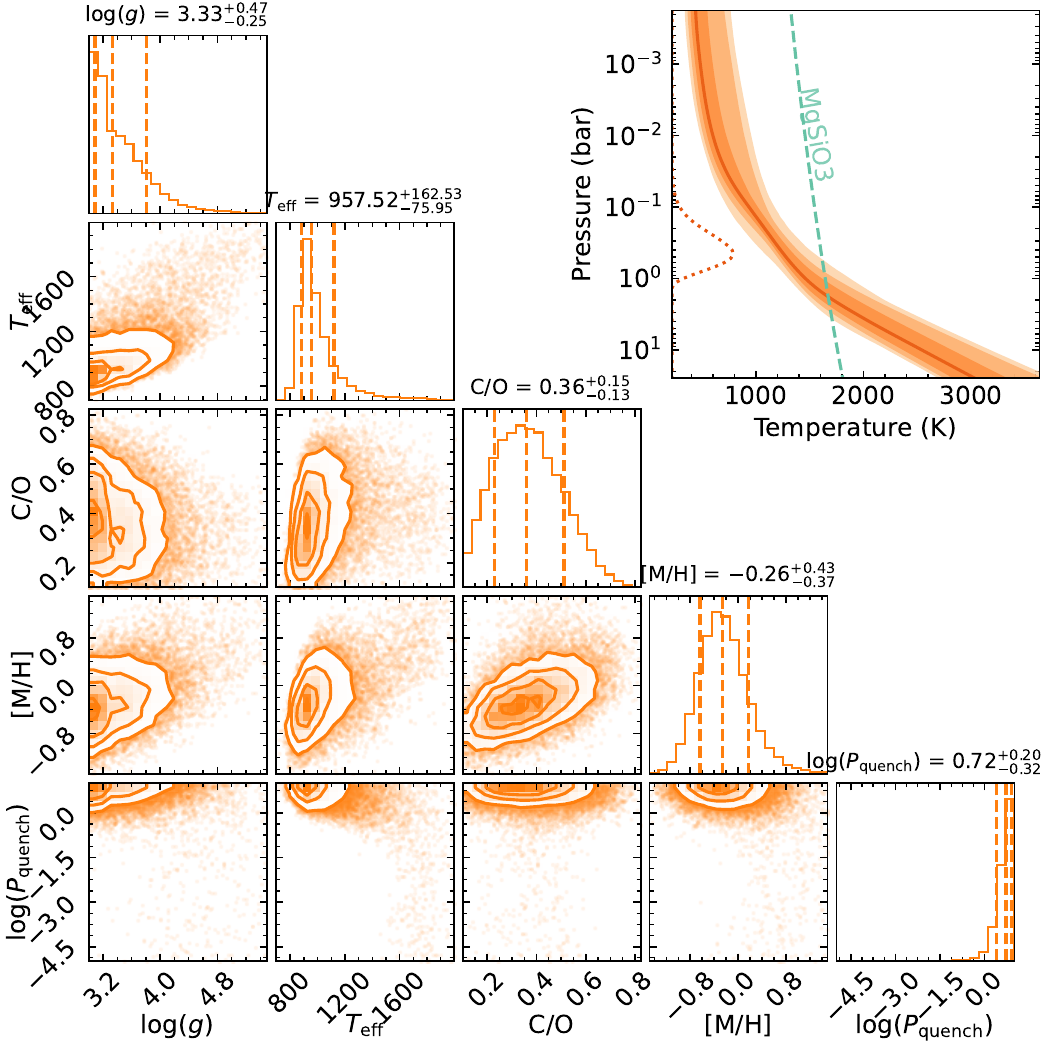}
    \caption{Posterior distribution of retrieved parameters for YSES~1~c, including surface gravity, 
    effective temperature, C/O ratio, metallicity, and quenching pressure. The vertical dashed lines denote the medium and 1$\sigma$ intervals.
    The upper right panel shows the retrieved 1$\sigma$, 2$\sigma$, and 3$\sigma$ envelope of the T-P profile. 
    The dotted line is the flux-weighted emission contribution. The dashed line is the condensation curve of \enstatite~clouds, 
    located below the region of the main emission contribution.
    \label{fig:1c_corner}}
\end{figure}

We detect \HTWOO~and CO in the atmosphere of YSES~1~c, as shown in the
cross-correlation functions in Fig.~\ref{fig:1c_ccf}. 
The CCFs suggest an S/N of 7, 4, and 7.8 for \HTWOO, CO, and \HTWOO+CO, respectively.
Using the leave-one-out Bayesian model comparison analyses, we quantify the detection significance of \HTWOO~and CO 
to be 7.3 and 5.7 $\sigma$, respectively.
The disequilibrium chemistry model results in a C/O$=0.36\pm0.14$ and a [M/H]$=-0.26\pm0.40$ (see Fig.~\ref{fig:1c_corner}). Again, the metallicity constraint is likely model-dependent. 
The C/O is broadly consistent with the solar value of 0.59 \citep{AsplundEtAl2021}, while peaking at slightly lower values.
However, it is not clear whether the C/O ratio of YSES~1~c is indeed sub-solar or still consistent with solar. 

The free chemistry model retrieves consistent results with the disequilibrium chemistry model (Table~\ref{tab:1c_params}).
The non-detection of \methane~places a stringent constraint on its volume mixing ratio
(3$\sigma$ upper limit of $10^{-5}$). 
This is consistent with strong quenching retrieved by the disequilibrium chemistry model,
which constrains a quenching pressure of $P_\mathrm{quench}>3$ bar (1$\sigma$).
Following \cite{XuanEtAl2022}, we convert the quench pressure to the vertical eddy diffusion coefficient
$K_\mathrm{zz}$ by equating the chemical timescale of CO-\methane\ reaction from \cite{ZahnleMarley2014} 
to the vertical mixing timescale $\tau_\mathrm{mix}$:
\begin{equation}
    \tau_\mathrm{mix} = \frac{L^2}{K_\mathrm{zz}} = \frac{\alpha^2H^2}{K_\mathrm{zz}} 
    = \frac{\alpha^2}{K_\mathrm{zz}} \bigg(\frac{k_B T}{\mu m_H g}\bigg)^2,
\end{equation}
where $L$ is the mixing length, which can be represented as the atmospheric scale height $H$ 
multiplied by a scaling factor $\alpha$. Adopting $\alpha=1$, we obtained a $K_\mathrm{zz} > 10^{10}$ cm$^2$\,s$^{-1}$ (1$\sigma$).
This value is close to the upper limit $\sim3\times10^{10}$ cm$^2$\,s$^{-1}$ predicted by the mixing length theory \citep{ZahnleMarley2014}.
Such strong quenching has been inferred in many super-Jovian companions and brown dwarfs 
near the L/T transition \citep[e.g.][]{KonopackyEtAl2013, BarmanEtAl2015, MolliereEtAl2020, XuanEtAl2022, deRegtEtAl2024, ZhangEtAl2024}, 
suggesting that disequilibrium chemistry plays an essential role in their atmospheres.

\begin{figure*}[t!]
    \gridline{\fig{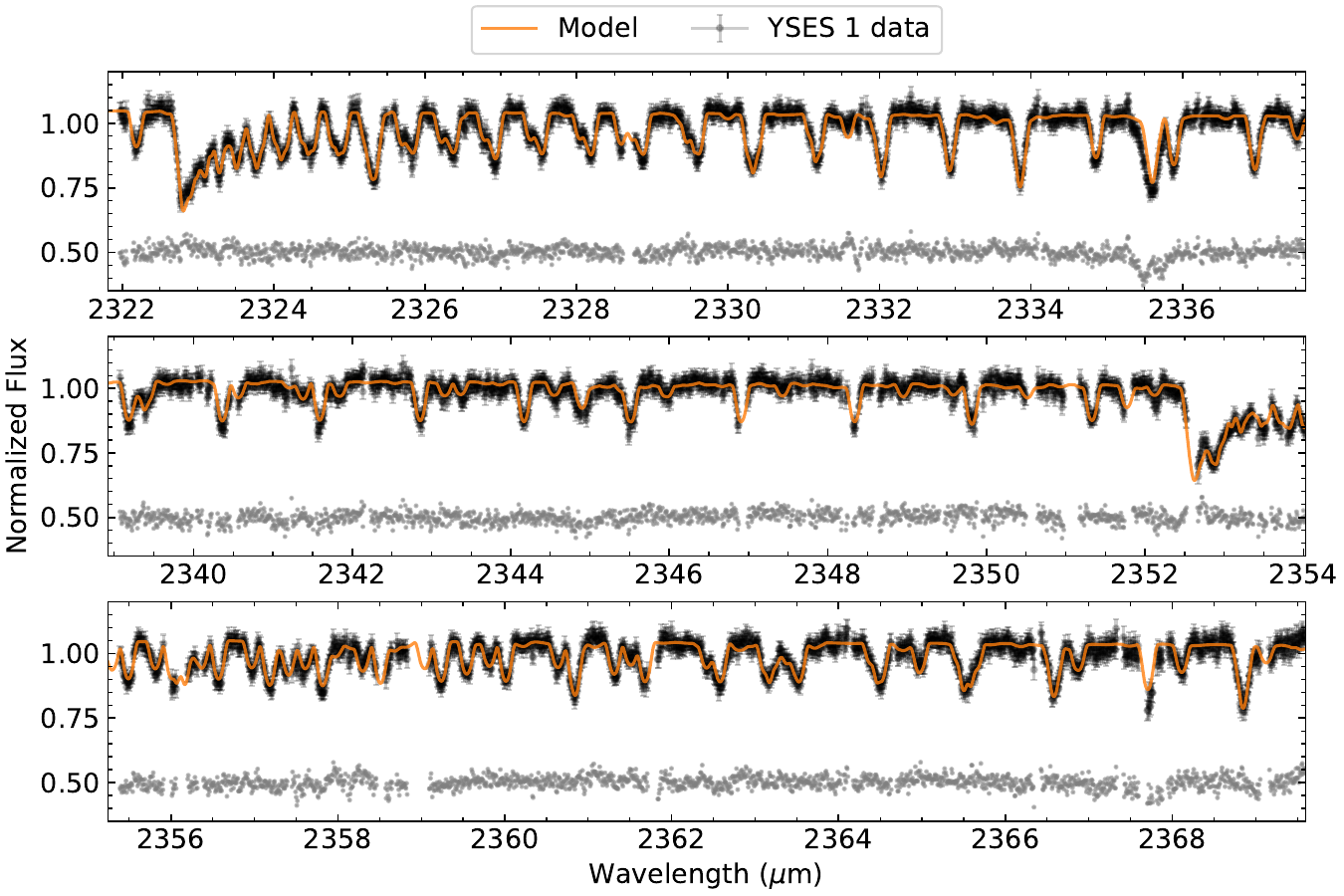}{0.65\linewidth}{}
    \fig{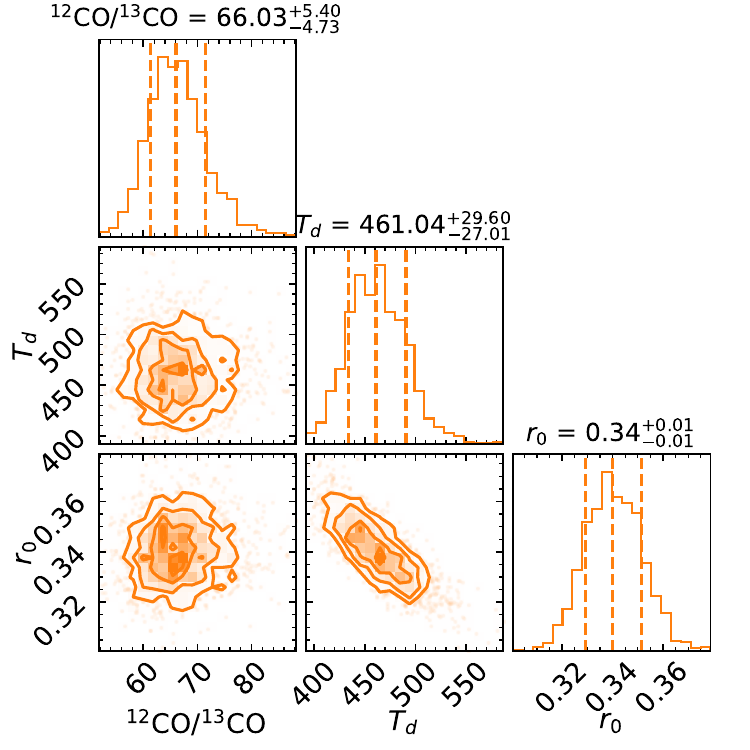}{0.35\textwidth}{}
    }
    \caption{Spectral analysis of the YSES 1 primary to retrieve the stellar carbon isotope ratio. Left panels show a portion of the CRIRES$^+$ spectra in black, the best fit PHOENIX stellar model in orange, and the residuals in gray (with an offset by 0.5). Right panel shows the posterior distribution of the retrieved \COmain/\COiso~ratio, disk temperature $T_d$, and the veiling  amplitude $r_0$. 
    \label{fig:1A_spec}}
\end{figure*}

\subsubsection{Temperature and clouds}

The effective temperature is constrained to be \Teff=$958_{-76}^{+163}$ K,
consistent with $1240_{-170}^{+160}$ K derived from broadband photometry \citep{BohnEtAl2020a} within $\sim$1$\sigma$. We also test the flexible T-P parameterization (see Section \ref{sec:forward-tp}) and find consistent retrieval results.
The derived low surface gravity is broadly in line with the planet mass and its red color. 
Although the red color of YSES~1~c indicates a cloudy atmosphere, our observations with CRIRES$^+$ provide little constraints on clouds. 
Given such temperature structure, the \enstatite~clouds are expected to condense slightly below the pressure range that the K-band observations are able to probe (see the T-P profile in Fig.~\ref{fig:1c_corner}).
Therefore, our observations likely do not probe the region that is most significantly affected by silicate cloud opacity, and therefore provide weak constraints on its cloud properties.
Turning off clouds in the retrieval model results in a Bayes factor of approximately 1, indicating that the data are equally well matched by cloudy and cloud-free models.
We tested different cloud species such as Na$_2$S and KCl clouds, but found no appreciable effect on the retrievals.
The low S/N of the observation also reduces our sensitivity to clouds.
Observations at other wavelengths, such as the JH band with stronger cloud opacities and the mid-infrared with silicate features \citep{MilesEtAl2023}, have a better chance to unveil the nature of clouds. The JWST program GO 2044 will be able to answer this question.

\subsection{\COmain/\COiso~ratio of the primary YSES~1} \label{sec:stellar_iso}

To strengthen the constraining power of carbon isotope ratios, we analyze the high-resolution spectra of the primary YSES~1 in our CRIRES$^+$ dataset to measure the stellar \COmain/\COiso~ratio and compare it to that of the companion.
We perform grid fitting to the stellar spectra observed in two nights using a custom PHOENIX model grid \citep{HusserEtAl2013} with varying \COmain/\COiso~ratios ranging from 16 to 301 with a step of 15 (GonzalezPicos et al. \textit{in prep.}).
As the primary has similar properties as the Sun (see Table \ref{tab:system}), we adopt \Teff=4600 K, $\log g$=4.5, and a solar metallicity for the stellar models. 
We model the possible veiling effect of a circumstellar disk on the spectra by considering the continuum emission contribution from a blackbody with a temperature $T_d$ and a veiling parameter $r_k$ \citep{GonzalezPicosEtAl2024}:
\begin{align}
    F_{\lambda} = \phi\big(F_{\lambda, \text{PHOENIX}} + r_k F_{\lambda}(T_{\text{d}})\big),
\end{align}
where $r_k$ represents the amplitude of the disk flux relative to the continuum of the photospheric emission at the reddest wavelength ($\lambda 2472$ nm) and $T_d$ is the effective temperature of the disk used in the single-blackbody disk model $F_{\lambda}(T_{\text{d}}) \sim B_{\lambda}(T_{\text{d}})$. Following \citealt{RuffioEtAl2019} (also \citealt{deRegtEtAl2024,GonzalezPicosEtAl2024}), we incorporate flux scaling factors $\phi$ calculated at every likelihood evaluation for each order-detector pair to allow offsets due to instrumental effects.
In addition to the two veiling parameters, the free parameters also include the carbon isotope ratio \COmain/\COiso, the radial velocity, and $v\sin i$ of the star.
The best-fit model to the data and posteriors of free parameters are shown in Fig.~\ref{fig:1A_spec}. 
We constrain the stellar \COmain/\COiso~abundance ratio to be $66\pm5$, consistent with the present-day ISM value of $\sim68$. We note that the uncertainty on the stellar \COmain/\COiso~only accounts for the statistical uncertainty, which should be regarded as a lower limit.

\subsection{Rotation and RV}

\begin{figure}[t]
    \includegraphics[width=\linewidth]{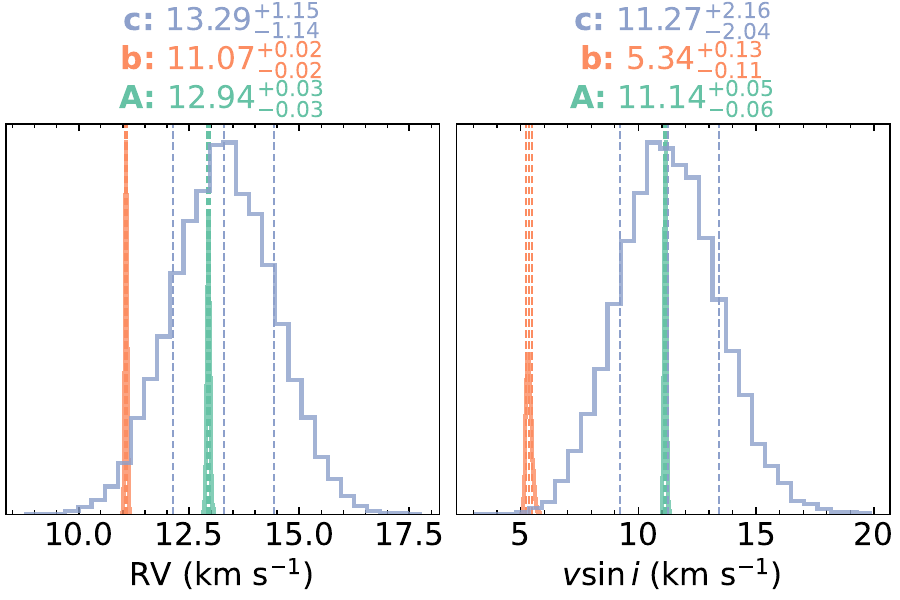}
    \caption{RV and $v\sin i$ of the primary YSES~1 and its companions b and c as constrained by CRIRES$^+$ observations.
    The vertical dashed lines represent the median and 1$\sigma$ values of the posterior distribution.
    \label{fig:rv_vsini}}
\end{figure}

We retrieve the rotation and radial velocity of the system as shown in Fig.~\ref{fig:rv_vsini}.
We measure an RV of $11.07 \pm 0.03$ \kms~for YSES~1~b and $ 13.3\pm 1.1$ \kms~for YSES~1~c. 
The primary has an RV of $ 12.94 \pm 0.03$ \kms~and a $v\sin i = 11.14\pm 0.06$ \kms.
Therefore, the relative RV between the companions and primary are $-1.87 \pm 0.04$ \kms~and $0.4 \pm 1.1$ \kms~for b and c, respectively.
These relative RVs can be included in joint orbital fitting with astrometric measurements 
from VLT/SPHERE and GRAVITY to refine the orbital constraints of the companions.

The projected rotation velocities of YSES~1~b and c are constrained to be $5.34 \pm 0.14$ and $11.3\pm 2.1$ \kms, respectively. 
Adopting isotropic distribution for the inclinations, we estimated the spin velocities of the two planets to be $\sim$6.2 and 14.4 \kms, 
which correspond to rotation velocity versus break-up velocity of $v/v_\mathrm{break}\sim0.05$ and 0.13.
We discuss the spin of YSES~1~b and c in the context of other super-Jovian companions in Section~\ref{sec:spin}.

\section{Discussion} \label{sec:discussion}

\begin{figure}[t!]
    \includegraphics[width=\linewidth]{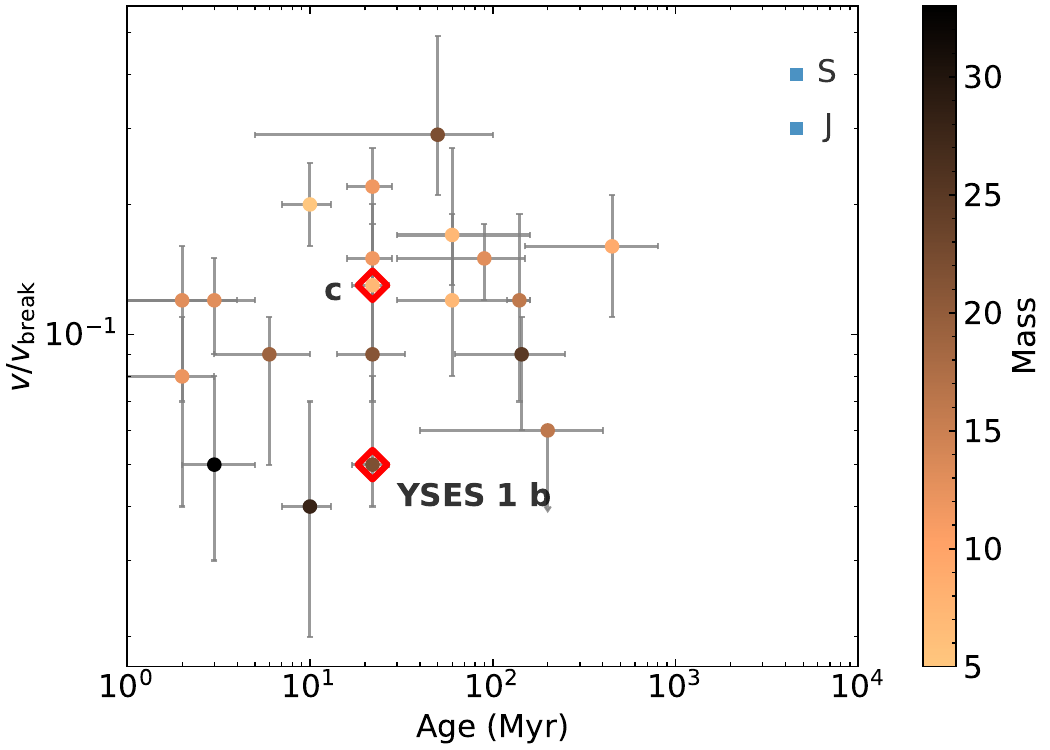}
    \caption{Rotation velocities (assuming isotropic inclination distribution) as a fraction of break-up velocities for super-Jovian companions.
    Our measurements of YSES~1~b and c are denoted with red diamonds.
    The measurements of other companions are adopted from the literature, including 
    \cite{BryanEtAl2020, WangEtAl2021, XuanEtAl2024a}. The data are color-coded by the companion masses.
    The blue data points show the spin of Jupiter and Saturn in the solar system.
    \label{fig:spin}}
\end{figure}

\begin{figure*}[t]
    \includegraphics[width=\linewidth]{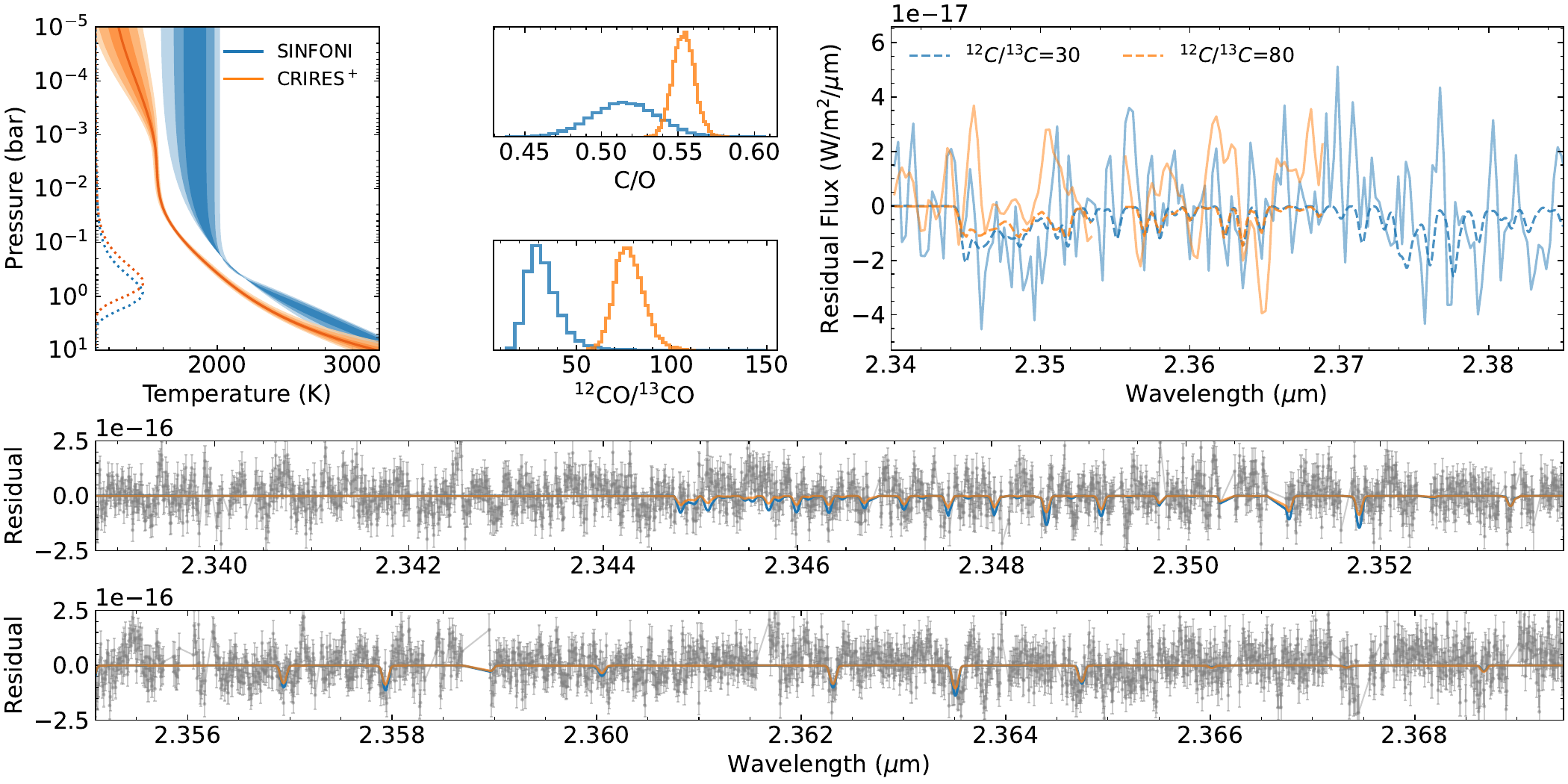}
    \caption{Comparison of the CRIRES$^+$ (orange) and SINFONI (blue) data and retrieval results on YSES~1~b.
    The left and middle panels show the posteriors of the T-P profile, C/O, and \COmain/\COiso~ratios.
    The right panel compares the observational residuals of the SINFONI data
    and the down-convolved CRIRES$^+$ data in solid lines and their best-fit \COiso~models in dashed lines.
    The two bottom panels show the observational residuals of the CRIRES$^+$ data like in Fig.~\ref{fig:1b_ccf}, overplotted with the retrieved best-fit \COiso~models from both CRIRES$^+$ (orange) and SINFONI (blue) data. 
    \label{fig:compare_sinfoni}}
\end{figure*}

\subsection{Trend of Rotation} \label{sec:spin}

Comparing the spin velocity of YSES~1~b and YSES~1~c, we note the $v\sin i$ of planet c is higher than that of b 
at the 3$\sigma$ level.
The distinct projected rotation velocities of super-Jovian companions in the same system are intriguing. 
This may indicate either different spin axis inclinations or different formation histories.
On the one hand, the two planets may have similar rotation rates, while the companion b has a small 
spin axis inclination that leads to a low projected spin velocity. This is consistent with the non-detection of 
polarized scattered light from the circumplanetary disk around YSES~1~b \citep{vanHolsteinEtAl2021}. 
On the other hand, the projected rotation velocities may reflect the actual spin rates of the two planets,
implying that they may have undergone different formation and evolution processes despite sharing the same 
natal environment. 
The long-lived circumplanetary disk (CPD) around the companion b may have might have helped to slow its spin rate via efficient magnetic braking  \citep{Batygin2018, GinzburgChiang2020, WangEtAl2021}. 
A similar trend of rotation rate with planet mass has been suggested for HR 8799 b and c \citep{WangEtAl2021}.
It is hypothesized that massive companions can effectively ionize the CPDs and spin down through interactions between magnetic fields and CPDs \citep{Batygin2018}.
In contrast, the CPD around planet c might have a shorter lifetime due to the larger orbital separation of 320 au,
or the magnetic braking is less efficient because of the lower planet mass.
We will be able to differentiate these two scenarios by measuring accretion signatures for companion c, for example, with the upcoming JWST observations.

Putting the rotation velocities into context, we show in Fig.~\ref{fig:spin} that YSES~1~b and c fit into the general trend 
of other super-Jovian companions from literature measurements.
Previous studies suggest that younger super-Jovian companions and brown dwarfs generally have lower rotation rates, 
which are expected to 
increase with age as their radii contract following angular momentum conservation \citep{BryanEtAl2020, VosEtAl2020}.
The rotation velocity (assuming isotropic inclination distribution) of YSES~1~b is among the lowest 
compared to super-Jovian companions at similar ages, while more comparable to that of younger companions ($<10$ Myr) 
such as GQ Lup b \citep{SchwarzEtAl2016} and HIP 79098 b \citep{XuanEtAl2024a}.
It is also interesting that these slow rotators are relatively massive ($>$15 \Mjup) compared to their low-mass 
counterparts ($<$10 \Mjup) such as $\beta$ Pic b \citep{SnellenEtAl2014, LandmanEtAl2024, ParkerEtAl2024}, 
HR 8799 bc \citep{WangEtAl2021}, and YSES~1~c.
This is consistent with the tentative trend between spin and mass \citep{WangEtAl2021}. 
However, measurements of a larger sample are needed to verify the trend.
Additionally, future rotation period measurements with light curves will be essential to break the 
degeneracy with inclinations and allow for comparisons of deprojected spin velocities.

\subsection{YSES~1~b with CRIRES$^+$ versus SINFONI} \label{sec:sinfoni}

\cite{ZhangEtAl2021a} has characterized the emission spectrum of YSES~1~b using medium-resolution 
spectroscopy with VLT/SINFONI ($\mathcal{R}\sim 4500$) and found a \COmain/\COiso=$31_{-10}^{+17}$.
Our CRIRES$^+$ observations suggest a carbon isotope ratio of $88\pm13$, which is 2.7$\sigma$ higher than the SINFONI measurement.
To investigate the discrepancy, we compare the retrieval results of the two datasets as shown in Fig.~\ref{fig:compare_sinfoni}.
In addition to the carbon isotope ratio, the retrieved T-P profiles differ significantly -- 
the SINFONI retrieval converged to a higher temperature ($>$2000 K)
combined with a smaller radius (1.7 \Rjup) to reproduce the same flux. 
It is not clear why the SINFONI data prefer such a combination. 
We test the SINFONI data using our new retrieval framework (see Section~\ref{sec:retrieval}) and find results similar to those of \cite{ZhangEtAl2021a}.
This rules out the possibility that the different T-P parameterizations may bias the retrieval results. 
Instead, it points to inconsistencies between the two datasets.

Focusing on the region of \COiso~absorption, we compare the observational residuals of the SINFONI data
and the down-convolved CRIRES$^+$ data in Fig.~\ref{fig:compare_sinfoni}. 
The corresponding best-fit \COiso~model spectra for different abundances are overplotted. 
The difference in isotope ratios is mainly 
reflected in the depth of the \COiso~bandhead absorption around $2.345-2.352$ \micron, 
where the CRIRES and SINFONI data show significant discrepancies. The SINFONI data show an apparent dip near the 
bandhead of \COiso~which likely drives the high abundance of \COiso~inferred from the data.
Meanwhile, the bandhead absorption feature is less prominent in the CRIRES observation.

If both observations are valid, they might be explained by atmospheric variability. Hypothetically, 
variable clouds or veiling may affect the inference of isotope ratios as \COmain~lines probe lower 
pressure levels than \COiso~lines due to stronger line opacities, and clouds or other opacity 
sources may block the features of minor isotopologue from the deeper atmosphere. However, this 
effect seems to be marginal according to our retrieval tests -- the enforced cloud model and 
veiling model did not alter the isotope ratio significantly. 

Alternatively, the observations might be biased because of systematic noise. 
Although we cannot rule out the SINFONI measurement, the CRIRES$^+$ result should be more reliable 
because of the higher spectral resolution and the detection significance of \COiso.
We estimate that the data quality of CRIRES$^+$ is better than SINFONI by:
\begin{equation}
    \frac{\mathrm{S/N}_\mathrm{CRIRES} \sqrt{R_\mathrm{CRIRES}}}{\mathrm{S/N}_\mathrm{SINFONI} \sqrt{R_\mathrm{SINFONI}}} \sim \frac{15\times\sqrt{116000}}{50\times\sqrt{4500}} \sim 1.5.
\end{equation}
The \COiso~detection in the high-resolution spectra relies on individual absorption lines at multiple locations in addition to the bandhead feature.
In contrast, since the medium-resolution observations cannot fully resolve the spectral lines, it may be possible that 
they are more susceptible to systematic noise that likely has frequencies comparable to the partially resolved 
molecular features (such as the \COiso~bandhead) in the planetary atmosphere.
The \COiso~bandhead region is particularly difficult for telluric correction due to blended \methane~and \HTWOO~lines.
The systematics may result from nuances in data reduction that we are unaware of, 
such as stellar background removal and telluric transmission correction. 
While the systematics do not affect the detection of isotopes, they may still bias the abundance constraints.
This underlines the challenge of measuring marginal spectral features and comparing different spectral resolutions.
More work is needed to assess how the data reduction affects retrieval results and properly 
model correlated noise in medium-resolution data.

\subsection{Implication for planet formation}

The primary star YSES~1 is a young solar analog. GAIA DR3 suggests [Fe/H]=$-0.07\pm 0.01$ \citep{GaiaCollaborationEtAl2021}. 
For the following discussion, we assume the stellar chemical abundances to be solar-like.
Our retrieval analyses infer the C/O $=0.57 \pm 0.01$ in YSES~1~b
to be consistent with the stellar value.
Although the metallicity constraint is correlated with the surface gravity, we can safely reject high-$\log g$ and metal-rich solutions, based on the companion mass. Therefore, its metallicity is likely close to solar.
This star-like composition suggests that YSES~1~b likely formed via gravitational instability or core accretion beyond the CO iceline
where most volatiles freeze out onto dust grains, so the solid reservoir represents the proto-stellar composition. 
YSES~1~b shows a \COmain/\COiso=$88\pm13$, similar to the terrestrial value of $\sim89$ \citep{AndersGrevesse1989}, and consistent with the host-star carbon isotope ratio of $66\pm5$ within 1.6$\sigma$. 
If the marginally higher \COmain/\COiso~ratio in the companion is real, this may indicate the accretion of $^{13}$C-depleted gas from the disk, which is plausible as a result of the isotope selective photodissociation and ion exchange reaction \citep{Visser2009, MiotelloEtAl2014, BerginEtAl2024}. Hence, a deviation of the \COmain/\COiso~ratio from the stellar value would be an essential probe for distinguishing core accretion from gravitational instability.
Better constraints on the isotope ratio and metallicity will be critical to break the degeneracy in formation scenarios. The combination of high-resolution capability from the ground and the broad wavelength coverage from JWST will allow us to achieve this goal.

For YSES~1~c, the C/O $=0.36 \pm 0.14$ is subsolar or solar.
The low S/N detection limits 
our ability to draw tight constraints on the chemical abundances.
If the C/O ratio of YSES~1~c is stellar, it is consistent with in-situ formation like YSES~1~b via either core accretion or 
gravitational instability. 
On the other hand, if the C/O ratio in YSES~1~c is substellar, it indicates that the planet likely formed within the CO iceline, typically at 5 - 30 au for solar-type stars \citep{QiEtAl2013}.
The planet may have incorporated oxygen-rich
solids within the CO iceline or have accreted water vapor close to the water iceline. 
Considering the current location of planet c at $\sim$320 au, it then requires scattering events to relocate the planet. 
This would leave imprints in the orbital dynamics of both planets, which can be tested with future astrometry measurements. 
The substellar C/O ratio in YSES~1~c would also be inconsistent with YSES~1~b, implying different formation histories for the two companions.
With a lower planet mass of 7 \Mjup, YSES~1~c is more likely to have a bottom-up formation pathway and a deviation from the stellar composition.

Multi-planet systems like YSES 1 are valuable for understanding gas giant formation. Comparing chemical abundances in the atmospheres of both companions and the system's dynamical properties provides unprecedented details for tracing its formation history.
The upcoming JWST characterization of the system will deliver tighter constraints on the compositions of both companions. Constraints on the system architecture, such as orbits, spins, and alignments, using VLT/GRAVITY and CRIRES$^+$ provide complementary information, 
allowing us to test the hypothesis of the ex-situ formation of planet c and the system evolution as a whole.

Similar high-resolution spectroscopy analyses by \cite{XuanEtAl2024a} revealed a broad solar composition 
for a sample of 10-30 \Mjup~companions, while other super-Jovian companions with masses below 10 \Mjup, 
$\beta$ Pic b \citep{GRAVITYCollaborationEtAl2020, LandmanEtAl2024}, 
HR 8799 bcde \citep{KonopackyEtAl2013, MolliereEtAl2020, RuffioEtAl2021, WangEtAl2022, NasedkinEtAl2024}, 
51 Eri b \citep{Brown-SevillaEtAl2023,WhitefordEtAl2023}, 
AF Lep b \citep{ZhangEtAl2023, Palma-BifaniEtAl2024},
and YSES~1~c, 
display a larger diversity in their chemical composition. This may imply a transition in
formation pathways near the mass range of $5-10$ \Mjup~\citep{HochEtAl2023}. 
It agrees with the trend in companions' occurrence and orbital architecture that the low-mass super-Jovian companions ($<$10 \Mjup) 
and high-mass brown dwarfs may represent different formation channels \citep{NielsenEtAl2019, BowlerEtAl2020, ViganEtAl2021}.
The host-star metallicity trend with companion mass also features a transition around the same mass range \citep{Schlaufman2018}.
Charting the chemical diagram of low-mass super-Jovian companions
will allow us to determine whether a clear mass boundary separates this population from brown dwarfs 
or whether it represents the mixed outcomes of different formation mechanisms.

\section{Conclusion}\label{sec:conclusion}

We present the high-resolution spectroscopic characterization of the unique system
YSES~1 with two wide-orbit super-Jovian companions using the upgraded VLT/CRIRES$^+$ in K-band. 
We achieve a high S/N of $\sim$15 per pixel at $\mathcal{R}\sim 100,000$ for YSES~1~b with 5.2 hours of integration.
The observations also lead to atmospheric detection of the faint planet YSES~1~c ($K_\mathrm{mag}>18$).
We use free retrieval analyses to constrain the chemical and isotope abundances, temperature structure,
projected rotation velocity ($v\sin i$), and radial velocity (RV). 

\begin{itemize}
    \item We confirm the previous detection of \COiso~in YSES~1~b \citep{ZhangEtAl2021a} at a higher significance of 12.6$\sigma$.  
    The observations also reveal HF at 11.7$\sigma$.
    We constrain a C/O$=0.57 \pm 0.01$ and \COmain/\COiso=$88\pm13$ (1$\sigma$ confidence interval), consistent with the primary's isotope ratio $66\pm5$ as measured with the same dataset.
    This may suggest a formation via gravitational instability or core accretion beyond the CO iceline. 
    We note the retrieved shallow (isothermal) temperature profile compared to self-consistent models 
    and the model dependency of metallicity. 
    \item For YSES~1~c, we detect \HTWOO~and CO at 7.3 and $5.7\sigma$, respectively. The non-detection of \methane~results in 
    a quenching pressure higher than 3 bar (1$\sigma$), suggesting
    disequilibrium chemistry in this L/T transition atmosphere.
    Our observation is not sensitive to clouds because of the limited S/N and the clouds underneath the photosphere. 
    We constrain a subsolar or solar C/O of $0.36 \pm 0.15$, potentially indicating accretion of oxygen-rich 
    solids between \HTWOO~and CO icelines. Higher S/N observations are needed to constrain it better.
    \item We constrain the $v\sin i$ of companion b and c to be $5.34 \pm 0.14$ and $11.3 \pm 2.1$ \kms, respectively.
     Considering their same age and birth environment, this may indicate distinct spin axis inclinations or 
     effective magnetic braking by the longlived circumplanetary disk around YSES~1~b.
     We also measured the companion-to-primary relative RVs, which can be used to refine the orbital constraints in future studies. 
\end{itemize}

The CRIRES$^+$ observations highlight the capability of high-resolution spectroscopy to constrain various properties of 
spatially resolved exoplanets and push the limit to faint objects. 
YSES~1 represents an intriguing system for comparative studies of super-Jovian  companions and linking present atmospheres to formation histories.
Future astrometry and spectroscopy studies at wide wavelength ranges will be vital to unveil its origin.

\begin{acknowledgements}

    Y.Z. is thankful for the support from the Heising-Simons Foundation 51 Pegasi b Fellowship (grant \#2023-4298).
    Based on observations collected at the European Organisation for Astronomical Research in the Southern Hemisphere under ESO programme(s) 1110.C-4264(F).
    The computation was carried out on the Caltech High-Performance Cluster.
\end{acknowledgements}

\vspace{5mm}
\facilities{VLT/CRIRES$^+$}

\software{
\texttt{numpy}~\citep{HarrisEtAl2020},
\texttt{scipy}~\citep{VirtanenEtAl2020},
\texttt{matplotlib}~\citep{Hunter2007},
\texttt{astropy}~\citep{AstropyCollaboration2013, AstropyCollaborationEtAl2018, AstropyCollaboration2022},
\texttt{petitRADTRANS}~\citep{MolliereEtAl2019}, 
\texttt{PyMultiNest} \citep{BuchnerEtAl2014},
\texttt{corner}~\citep{Foreman-Mackey2016},
}

\appendix

\section{\texttt{excalibuhr} pipeline} \label{app:pipeline}

\texttt{excalibuhr} is an end-to-end pipeline that can extract and calibrate the 
high-resolution spectroscopic observations with VLT/CRIRES$^+$.
It follows the general calibration steps in the ESO's pipeline 
CR2RES\footnote{\url{https://www.eso.org/sci/software/pipelines/cr2res/cr2res-pipe-recipes.html}}, 
and \cite{HolmbergMadhusudhan2022}, while using customized implementation written in Python. 
The main steps of \texttt{excalibuhr} pipeline are summarized as follows.


\begin{itemize}
    \item \textbf{Calibration frames.} The raw dark and flat frames are median combined, 
    which results in master dark frames, flat, readout noise, and bad pixel map. 
    The flat and lamp frames are corrected for the dark. We then identify the traces of
    illuminated spectral orders on the detector images of the master flat using 
    second-order polynomials. 
    In addition to the curvature in spectral orders,
    the slit is tilted on detector images due to instrument geometry. 
    We measure the slit curvature of each line in the Fabry-P\'erot Etalon (FPET) frame 
    using second-order polynomials. The varying curvature along the dispersion direction
    is captured by fitting second-order polynomials to the polynomial coefficients from the previous step as functions of wavelengths.
    These measurements of order and slit curvatures can be used to straighten 
    the detector images to align traces in both spatial and spectral directions.
    The FPET frame also aids in obtaining an initial wavelength solution.
    We map the locations of FPET lines on detector images to evenly spaced wavelength grids
    using second-order polynomials, which provide the relative spacing of wavelength solutions. 
    They are further calibrated against the Uranium-Neon lamp lines using cross-correlation to
    obtain the absolute wavelength solutions.
    To correct the pixel-to-pixel response and uneven illumination on detectors, 
    we extract the flux of the flat frame in each order and smooth it to obtain the blaze function,
    which is used to normalize the flat.

    \item \textbf{Science frames.} We subtract the AB nodding pairs to remove 
    dark current and sky background, and apply the normalized flat field calibration to science frames.
    Then, multiple exposures are mean-combined at each nodding position, A and B.
    We consider three main sources of uncertainties in detector images, including 
    readout noise, sky background, and photon noise. 
    The sky background noise is estimated from the corresponding science frame at the opposite nodding position.
    The photon noise from the target is added quadratically and is iteratively improved during the spectrum 
    extraction (see the following step).
    Using curvature measurements, we straighten the tilted spectral orders and slits in detector images
    by linear interpolation. The corrected 2D data are ready for spectrum extraction.


    \item \textbf{Optimal extraction.}
    We use the optimal extraction algorithm by \cite{Horne1986} as follows. 
    \begin{equation}
        f=\frac{\sum_x \mathbf{M} \mathbf{P}\mathbf{D} / \mathbf{V}}{\sum_x \mathbf{M} \mathbf{P}^2 / \mathbf{V}},
        \quad \operatorname{var}[f]=\frac{\sum_x \mathbf{M} \mathbf{P}}{\sum_x \mathbf{M} \mathbf{P}^2 / \mathbf{V}}.
    \end{equation}
    This method takes the sum of flux along the cross-dispersion dimension ($x$), weighted 
    by the empirical point spread function ($\mathbf{P}$) constructed from the 2D data ($\mathbf{D}$). 
    It iteratively rejects outliers ($\mathbf{M}$) caused by cosmic rays or bad pixels while 
    revising the variance ($\mathbf{V}$) and the extracted 1D spectrum.
    The extraction aperture is automatically determined based on the derived PSF
    to encompass 90\% of the flux and exclude the PSF wings with lower S/N.
    The blaze function is corrected in each order.

    \item \textbf{Wavelength solution.}
    In addition to the initial wavelength solution based on lamp observations, we carry out 
    a second-stage wavelength calibration against the telluric transmission spectral model.
    In each order, we apply a third-order polynomial to the initial wavelength solution 
    and optimize the correlation between an 
    observed, featureless spectrum of a telluric standard star and a template generated by 
    ESO's sky model calculator SkyCalc \citep{NollEtAl2012, JonesEtAl2013}.

    \item \textbf{Instrument response.}
    We estimate the instrument response by comparing observations of the telluric standard star
    to a PHOENIX stellar model \citep{HusserEtAl2013}. This provides the wavelength-dependent 
    instrument response that can be applied to the spectra of science targets.
    Since early-type standard-star spectra are firmly in the Rayleigh-Jeans regime, 
    potential uncertainties in its effective temperature have negligible effects on 
    the calibration of the spectral shape.

\end{itemize}

\section{Observed spectrum and best-fit models} \label{app:specs}

\begin{figure*}[ht]
    \includegraphics[width=\linewidth]{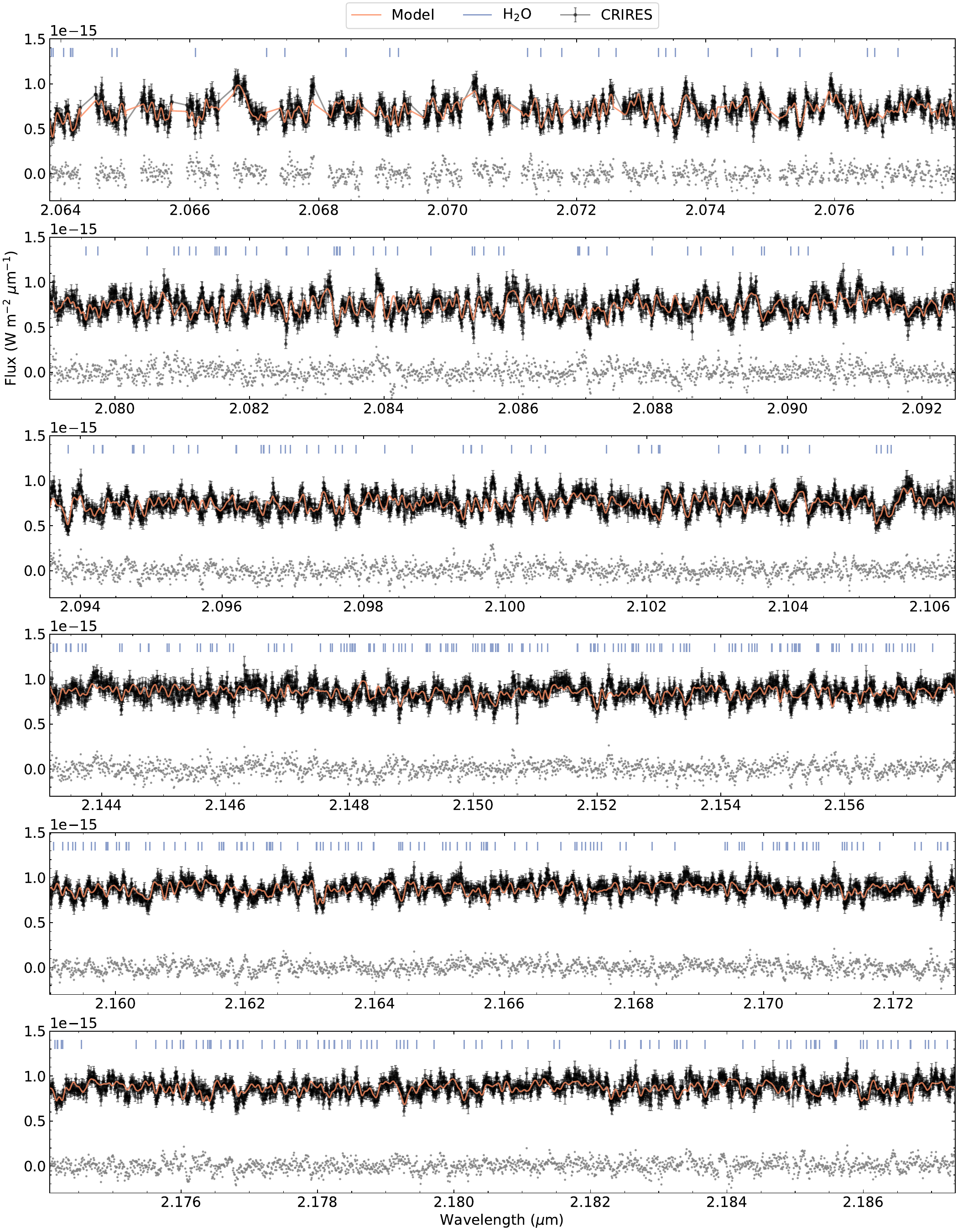}
    \caption{Similar as Fig.~\ref{fig:1b_spec} but for other spectral orders of CRIRES$^+$ observations of YSES~1~b. 
    \label{fig:1b_spec_1}}
\end{figure*}

\begin{figure*}[ht]
    \includegraphics[width=\linewidth]{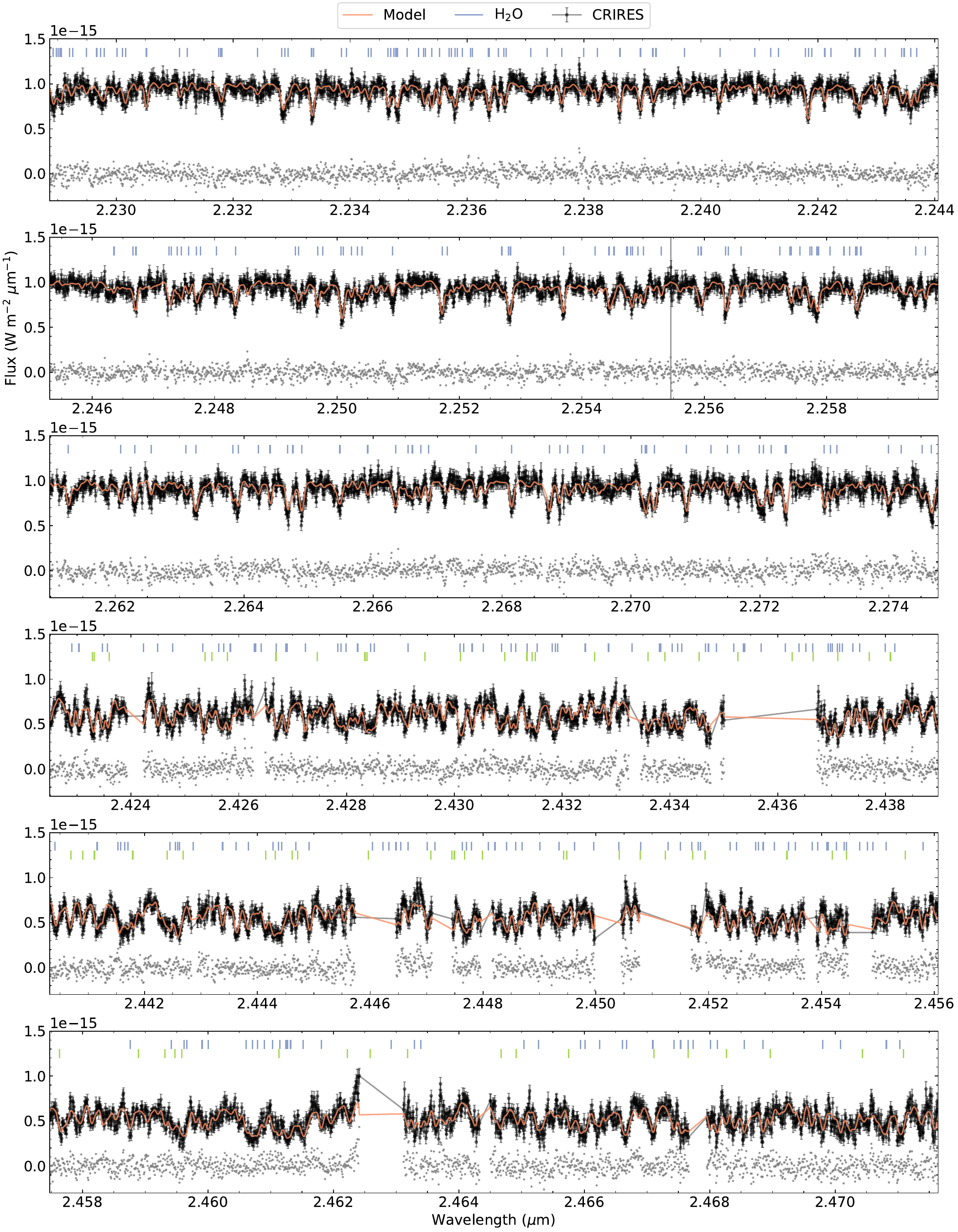}
    \caption{Fig.~\ref{fig:1b_spec_1} continued. 
    \label{fig:1b_spec_2}}
\end{figure*}

\begin{figure*}[ht]
    \includegraphics[width=\linewidth]{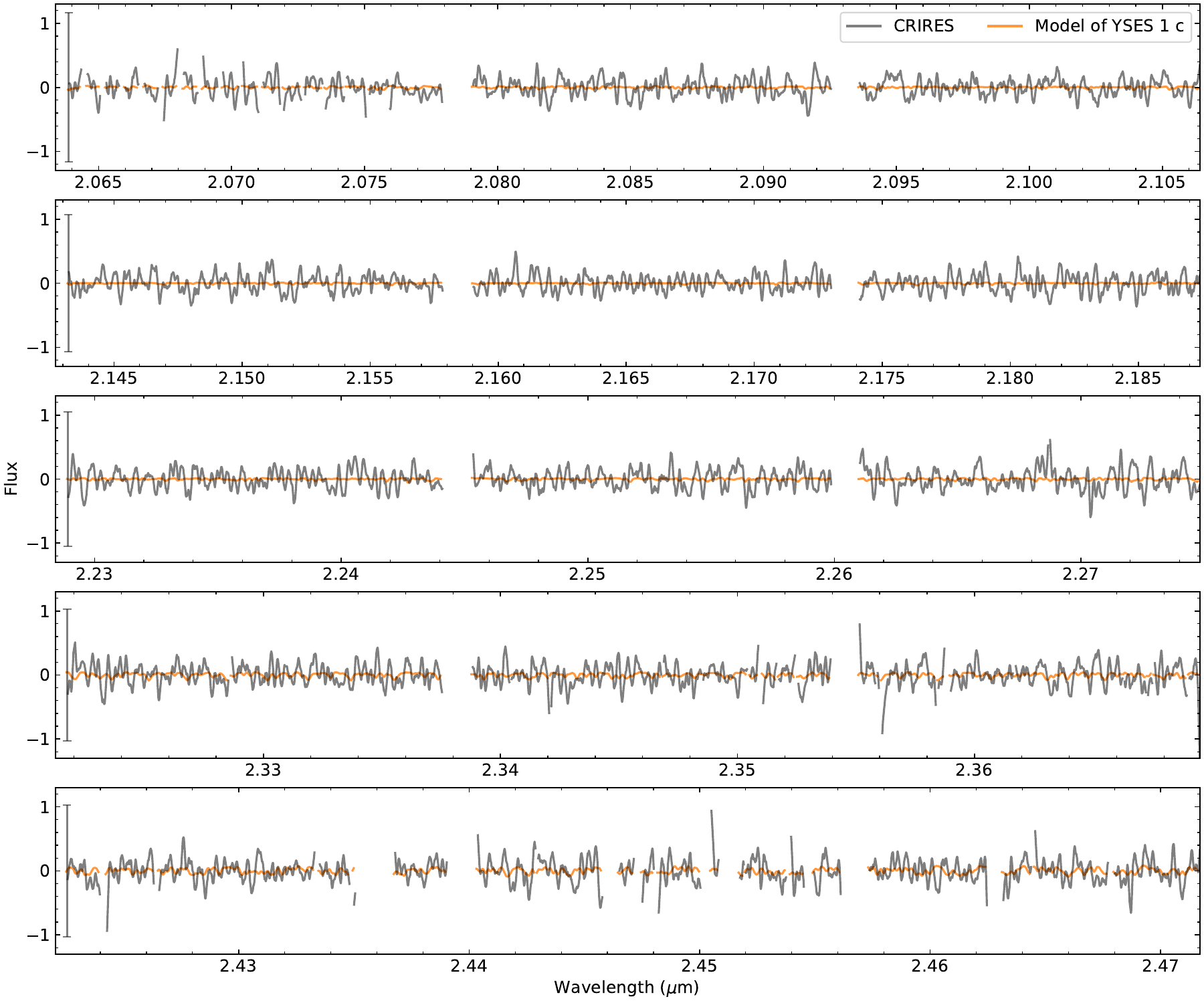}
    \caption{CRIRES$^+$ observations and best-fit model spectrum of YSES~1~c. 
    The observations in gray are convolved with the rotational broadening kernel for better visualization. The typical uncertainties of the data are shown in gray error bars on the left. The planetary signal is buried in the noise in individual channels.
    \label{fig:1c_spec}}
\end{figure*}

We show the full spectra (2.06 $-$ 2.47 $\mu$m) of CRIRES$^+$ observations on YSES 1 b and c with retrieved best-fit models in Fig.~\ref{fig:1b_spec_1}, \ref{fig:1b_spec_2}, and \ref{fig:1c_spec}.

\bibliography{manuscript}
\bibliographystyle{aasjournal}

\end{document}